\documentclass[10pt, conference, letterpaper]{IEEEtran}
\pdfoutput=1
\usepackage{cite}
\ifCLASSINFOpdf
   \usepackage[pdftex]{graphicx}
   \graphicspath{{../pdf/}{../jpeg/}}
   \DeclareGraphicsExtensions{.pdf,.jpeg,.png}
\else
\fi
\usepackage[cmex10]{amsmath}
\usepackage{array}
\usepackage[font=footnotesize]{subfig}
\usepackage{url}
\usepackage{balance}
\graphicspath{{Figures/}}
\usepackage{slashbox}
\usepackage{makecell}

\begin{document}
\title{Which Are You In A Photo?}

\author{\IEEEauthorblockN{Xing Zhang
%Yonggang Wen\IEEEauthorrefmark{2},
%Ming Liu\IEEEauthorrefmark{1}
}
%\IEEEauthorblockA{\IEEEauthorrefmark{1}School of Computer Science and Engineering\\
%University of Electronic Science and Technology of China,
%Chengdu, Sichuan, P.R China, 611731}
%\IEEEauthorblockA{\IEEEauthorrefmark{2}School of Computer Engineering\\
%Nanyang Technological University,
%Singapore, 639798}
Email: zhangsimba@gmail.com}


\maketitle

\begin{abstract}
%\boldmath
Automatic image tagging has been a long standing problem, it mainly relies on image recognition techniques of which the accuracy is still not satisfying. This paper attempts to explore out-of-band sensing base on the mobile phone to sense the people in a picture while the picture is being taken and create name tags on-the-fly. The major challenges pertain to two aspects - "Who" and "Which". (1) "Who": discriminating people who are in the picture from those that are not; (2) "Which": correlating each name tag with its corresponding people in the picture. We propose an accurate acoustic scheme applying on the mobile phones, which leverages the Doppler effect of sound wave to address these two challenges. As a proof of concept, we implement the scheme on 7 android phones and take pictures in various real-life scenarios with people positioning in different ways. Extensive experiments show that the accuracy of tag correlation is above 85\% within 3m for picturing.
\end{abstract}


% \fi
\IEEEpeerreviewmaketitle

\section{Introduction}

% Why doing this? First how related to daily life, then how related to research area
Identifying one or a number of certain people in a picture is a common need in the social area. Such as Facebook\cite{FacebookTag}, Google Picasa\cite{GoogleNameTag}, etc., they all have picture tagging services. However, the services are mostly operated manually all by users themselves. In fact, automatic image tagging has been studied for a long time in research area, while the fields of image processing and face recognition have made significant progress, it remains difficult to automatically label a given picture\cite{TagsenseTransaction}. Furthermore, with the explosion of digital pictures and the resulting growing requirement of image retrieval, the development of automatic image tagging is becoming urgent and crucial.

\begin{figure}[!t]
\centering
\includegraphics[width=0.45\textwidth]{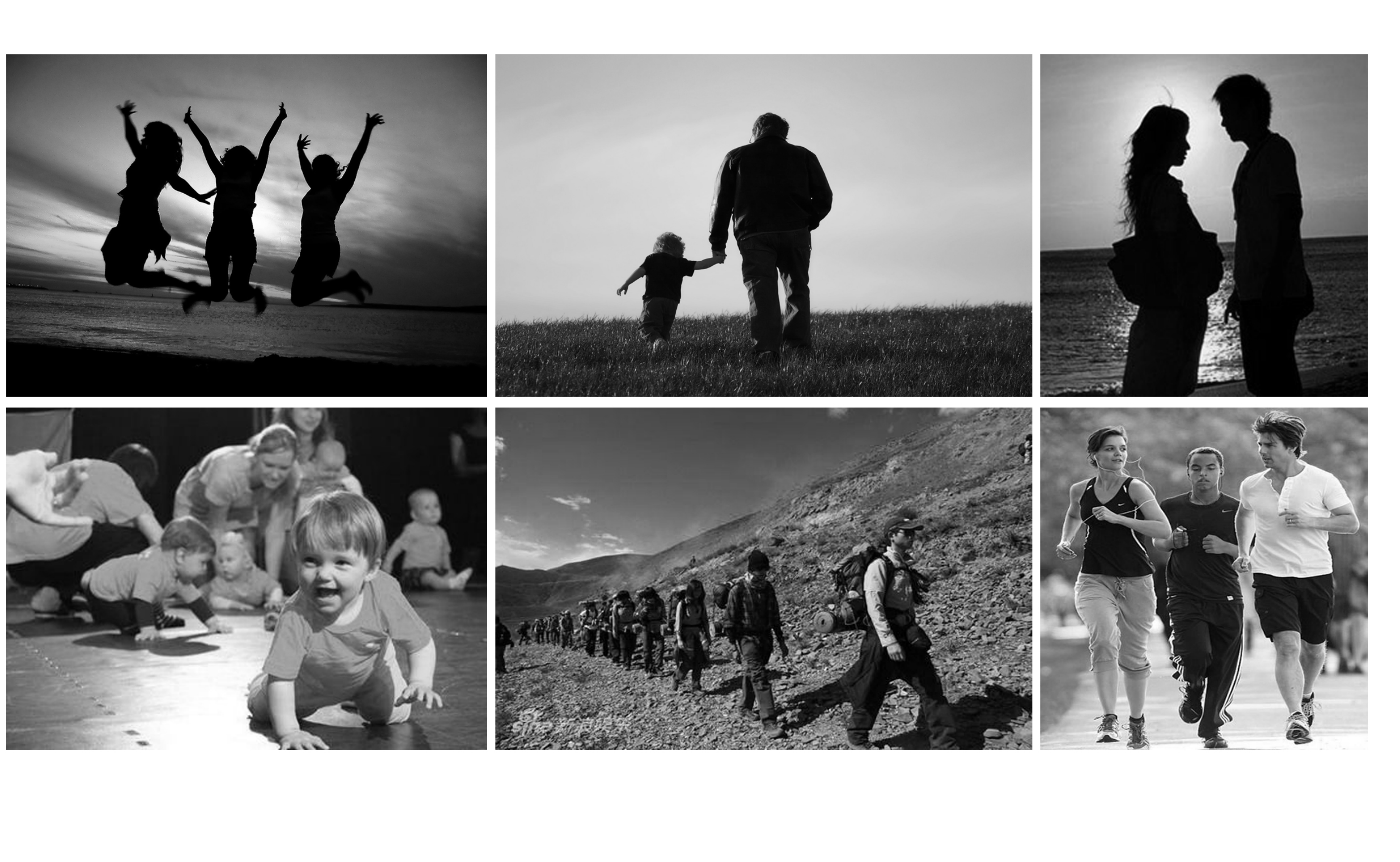}
\caption{Exceptions of the Tagsense: the first picture on the left top is shot when people jump up, which is against the "Pause Gesture" mechanism; people in the second and third pictures on the first row have their back and side toward the cameras, which are against the "Opposite Compass" mechanism; people in the last three pictures are crawling, walking and running, both people in the picture and not in the picture may act similarly, moreover, different depth of image field may lead to different motion analytical results which would bring severe difficulties to "Correlated Motion" mechanism.}
\label{fig::exceptionsoftagsense}
\end{figure}

% The core idea. How's current works? What's the key idea? Scenario?
Tagsense \cite{TagsenseTransaction} explored an alternative way to image tagging, using a multi-dimensional, out-of-band sensing of mobile phones. It tags photos with such a format "when-where-who-what", of which the content of "when" and "where" is obtained through timing and localization components, such as clock and GPS; the tag of "what" is obtained from motion analysis of sampled data from embedded inertial sensors, and the names of "who" in the picture come from a short-time interactive communication between neighboring users. Nevertheless, a neighboring user who is not in the picture may also respond with her name. The author then adopts three mechanisms to screen out the unrelated users: (1) Pause Gesture: people explicitly pose during picturing for most time, which means the related users would be stationary; (2) Opposite Compass: people in the picture often face to the camera, the compass readings of their mobile phone are opposite to the compass readings of the camera; (3) Correlated Motion: the movement of a certain people in the picture should be the same despite of the motion is described by image analysis or by accelerometer/compass readings. Although it is reported that the Tagsense has good precision and recall rate, we discreetly consider the three aforementioned mechanisms as not feasible in a number of common real-life photographic scenarios, as shown in Fig. \ref{fig::exceptionsoftagsense}.

This paper aims to give an alternative solution of "Who" and "Which", which are among the primary challenges in automatic image tagging\cite{TagsenseTransaction}. Different from the Tagsense based on the assumptions of human motion in front of cameras, the scheme we proposed is rooted in a physical law - the Doppler effect, which barely depends on the assumptions. The Doppler effect could be described as $f=\frac{c+v_R}{c-v_S}f_0$, where $f_0$ is the initial frequency of signal transmitted from a sender, $f$ is the frequency of the signal received by a receiver, $c$,$v_R$ and $v_S$ are the speed of the signal, of the receiver, and of the sender, respectively\cite{SoundWave}. Note that the $v_S$ is the speed of the sender moving in the direction towards the receiver, if the sender moves in a speed of $v$ at an angle $\theta$ to the receiver, then $v_S = v\cdot \cos\theta$. Same explanation could be made to $v_R$. It can be observed that when $c$, $f_0$ and $v$ is constant, $f$ is a monotonically decreasing function of $\theta$, and $\theta$ can be calculated if the $f$ is detected. This feature would be useful in addressing the aforementioned challenges which would be explained later.

The core idea of this paper is simple. Consider a scenario that each of Alice and her friends carries a mobile device which is embedded with a camera, a speaker, a microphone and a communication unit, such as the common off-the-shelf mobile phone. Alice is about to take a picture of her friends who have already positioned and posed, the moment before she presses the shutter, the speaker of her mobile device emits a short period of sound wave at a fixed frequency, and the mobile device moves in an orthogonal direction to the camera with a peak velocity $v$ simultaneously. According to the Doppler effect, the mobile devices carried by Alice's friends will receive the sound wave with a shifted frequency. Different relative positions make different frequency shifts. Based on this phenomenon, we could infer the relative positions of Alice's friends through sorting all the frequency shifts in order. At last, the names of Alice's friends will be tagged on the picture as per the relative positions.

% Why better? Challenges?

The problem bears resemblance to interaction and localization between neighboring mobile devices, where the Doppler effect is used by a few recent works, such as Spartacus\cite{Spartacus} and Doplink\cite{Doplink}. However, to our best knowledge it's the first time to utilize the Doppler effect in the context of automatic image tagging, and it raises some distinct research challenges. Two major challenges are summarized as two keywords: "Who" and "Which". (1) "Who": similar to the Spartacus\cite{Spartacus}, all the neighboring mobile devices will reply with the users' names before taking a picture, it would be difficult to screen out the people who are not in the picture without the help of visual comparison. (2) "Which": for the applications of image retrieval and social networks, it's important to correlate each tag with its corresponding features in a picture, which requires computational image analysis algorithms, like \cite{vision1}\cite{vision2}, which are not suitable for the energy-constrained mobile devices yet the accuracy remains around 47\% which is not satisfying\cite{vision2}.
% Any more clarification?
% Figure 1

\begin{figure}[!t]
\centering
\subfloat[Field Of View]{
\begin{minipage}[b]{0.22\textwidth}
\includegraphics[width=1\textwidth]{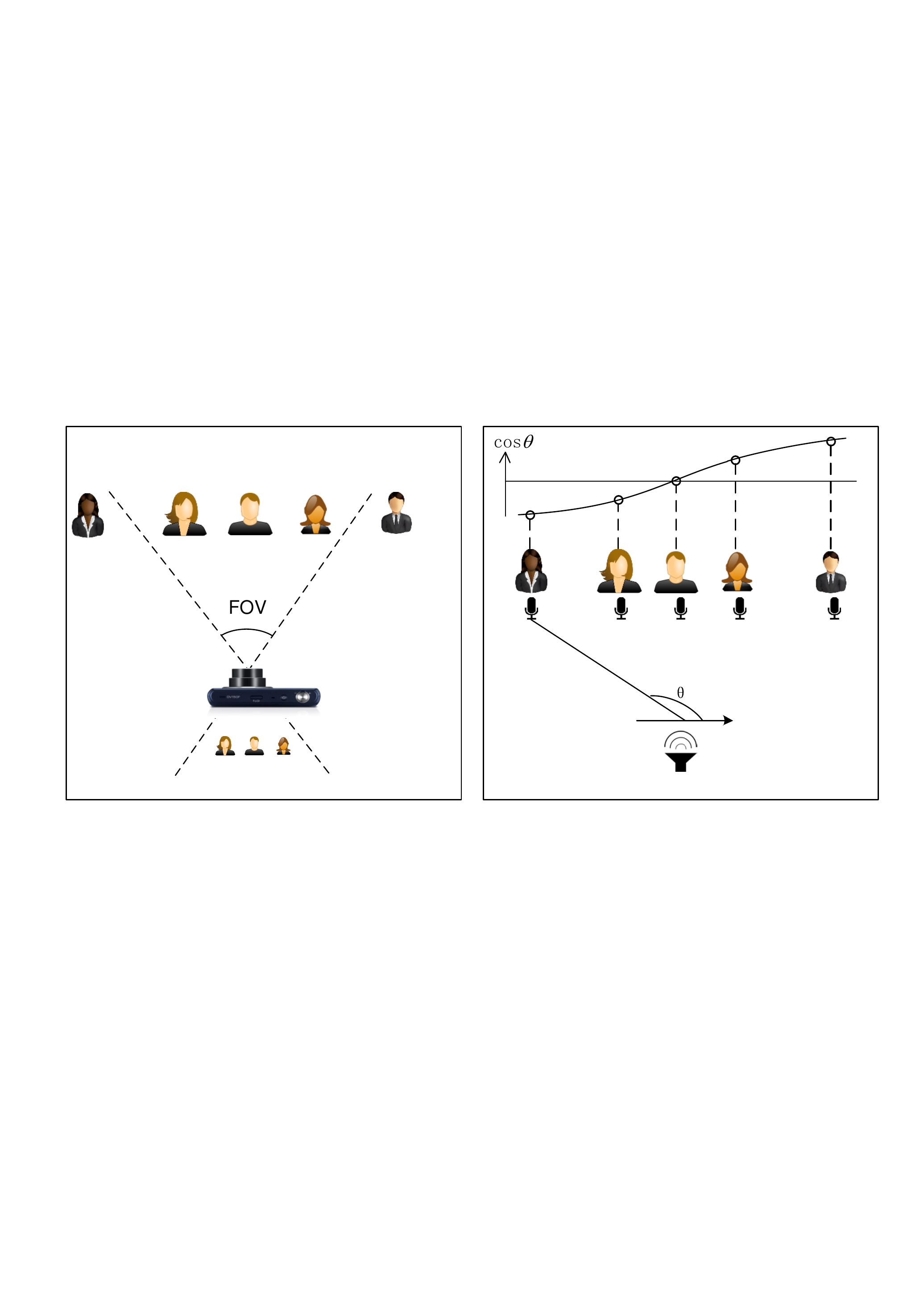}
\end{minipage}
\label{subfig::2-1}
}
\subfloat[Distinguishing frequency shifts]{
\begin{minipage}[b]{0.22\textwidth}
\includegraphics[width=1\textwidth]{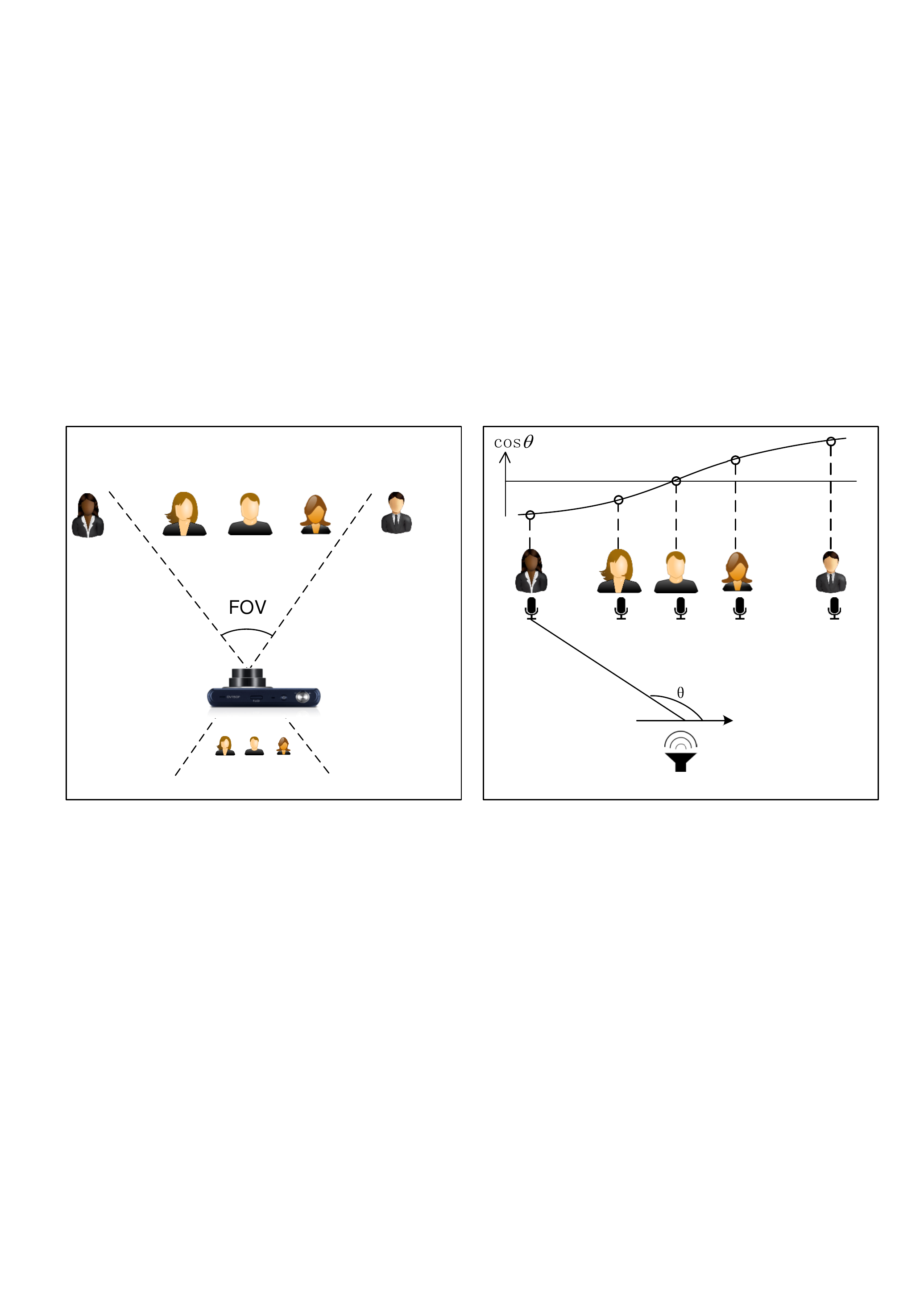}
\end{minipage}
\label{subfig::2-2}
}
\caption{Schematic of FOV and frequency shifts by the Doppler effect. (a) shows that people within the FOV will appear in the image, the ones outside the FOV will then be invisible; (b) illustrates that when the speaker of the sender moves to the right, the microphone of each receiver will receive a sound wave with frequency shifts according to the Doppler effect, different relative position correlated to different frequency shifts which are proportional to the $cos\theta$}
\label{fig::2}
\end{figure}

% How to address challenges?
In this paper, we propose an acoustic scheme applying on the off-the-shelf mobile phones to perform automatic name tagging on the pictures while they are being taken. For convenience, we call Alice's mobile phone with an audio emitting speaker as the sender, and other neighboring mobile phones with listening microphones as the receivers. In fact, the field of view (FOV) of a camera is the determinant factor of whether a person is captured by the camera, as shown in Fig. \ref{subfig::2-1}. According to the Doppler effect, the angle from each receiver to the sender can be calculated, therefore it would be easy to determine whether a user is in the picture through comparing the angle with the FOV. The analogous principle applies to the "Which" problem. As Fig. \ref{subfig::2-2} shows, if the speaker of the sender moves orthogonally to the orientation of the camera, each receiver will receive distinguishing frequency shifts inversely proportional to the angle $\theta$, the relative position of each user can be estimated as a consequence.

% Contributions
Our main contributions in this paper may be summarized as:
\begin{itemize}
\item \textbf{Proposed a Doppler effect based acoustic scheme to address the two challenging problems: "Who" and "Which" for automatic image tagging.} We present a novel acoustic technique based on the Doppler effect to enable accurate automatic name tagging, particularly we propose a multi-row recognition method to identify the relative position of each user in different rows if the users are positioned to multiple rows, like graduation class photography.
\item \textbf{Implemented the scheme on Android platform.} We implement the automatic name tagging scheme on common off-the-shelf Android mobile phones, the system runs as an application and needs not any extra equipment.
\item \textbf{Evaluated the system in real-life scenarios.} We test the performance of "Who" and "Which" in various real-life scenarios using 7 android mobile phones.
\end{itemize}

% Organization
The rest of this paper is followed by methodology in section \ref{sec::methodology}, implementation and evaluation in section \ref{sec::evaluation}, related works in \ref{sec::relatedworks} and conclusion in section \ref{sec::conclusion}.

\section{Methodology}\label{sec::methodology}

% This section gives a theoretical solution of automatic name tagging, mainly to address two challenges.
In this section, we will give theoretical details of how to address the aforementioned challenges under such a scenario. Assuming that Alice is taking a picture of her friends, each people has a mobile phone with a camera, a speaker, a microphone and a communication component embedded. We call the mobile phone of Alice as sender, and the mobile phones of her friends as receivers. The moment before Alice clicks the shutter for picturing with her mobile phone, the speaker of the sender emits a short period of audio tone at a known frequency $f_0$ meanwhile the speaker moves in a direction orthogonal to the orientation of the camera. The receivers will receive the audio and send back the user names and received frequency shifts to the sender. The sender will sort the received frequency shifts in order and tag the picture with user names accordingly.

% How to "Who"? How to ascertain the middle line?
\subsection{Who are in the picture?}
As mentioned before, the FOV of a camera is the determinant factor to judge who are in the picture, and it is a preset parameter which can be read from the rom of the mobile device. The challenge is how to obtain the angle of each receiver to the camera.
\begin{figure}
\centering
\subfloat[]{
\begin{minipage}[b]{0.22\textwidth}
\includegraphics[width=1\textwidth]{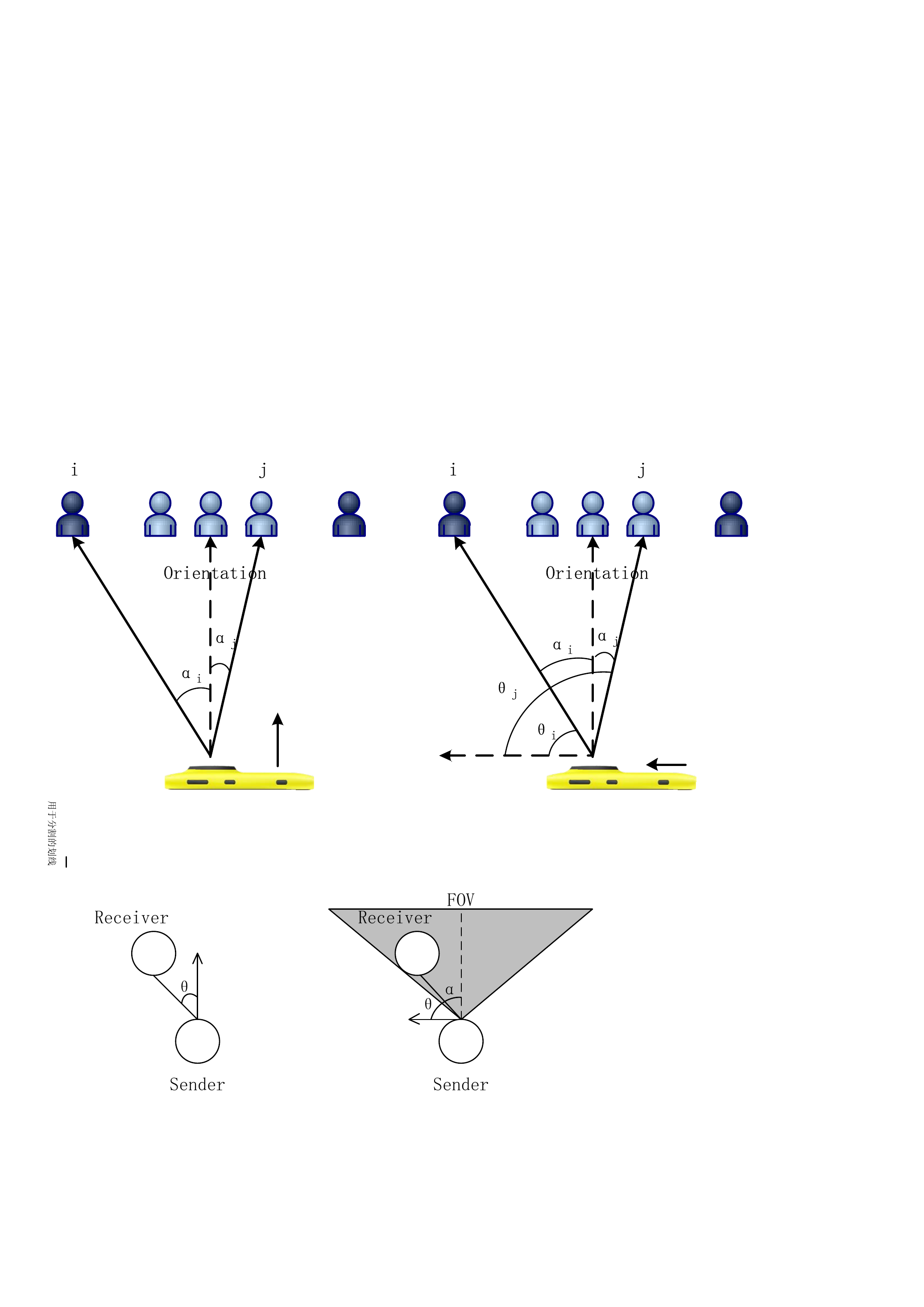}
\end{minipage}
\label{subfig::3-1}
}
\subfloat[]{
\begin{minipage}[b]{0.22\textwidth}
\includegraphics[width=1\textwidth]{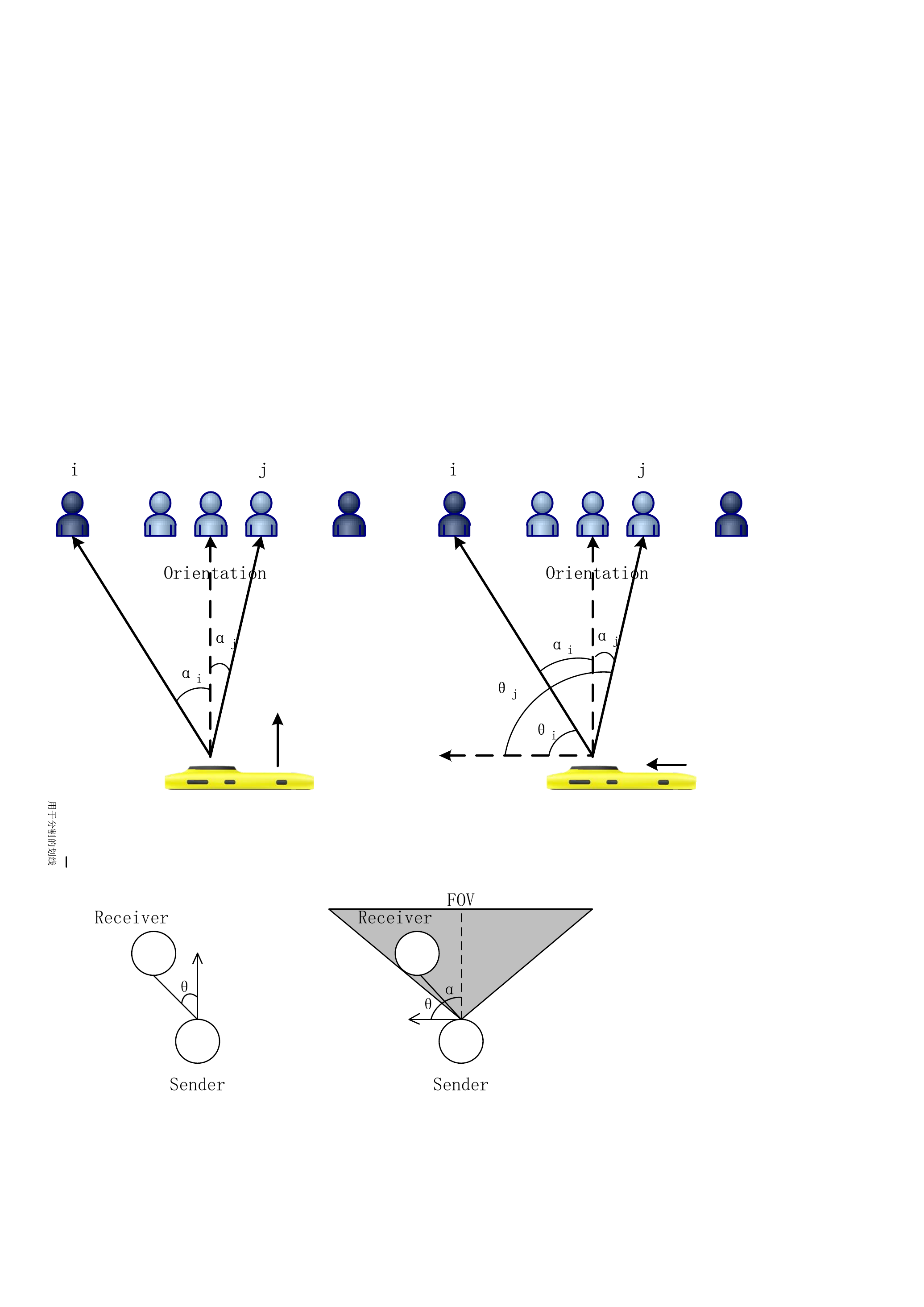}
\end{minipage}
\label{subfig::3-2}
}
\caption{The required angle for FOV determination. (a) the sender is moving towards the receiver, the $\theta$ is the required angle; (b) the sender is moving orthogonally to FOV area, the $\alpha$ is the required angle.}
\label{fig::3}
\end{figure}

The formulation of the Doppler effect $f=\frac{c+v_R}{c-v_S}f_0$ assumes that the sender is either directly approaching or receding from the receiver. If the sender approaches a receiver at an angle $\theta$ in a speed $v_S$, then the Doppler effect could be described as $f=\frac{c}{c-v_S\cdot\cos\theta}f_0$, assuming the receiver is stationary or moving vertically like jumping. Obviously, the $\theta$ is the angle required to determine whether the receiver falls in the FOV as Fig. \ref{subfig::3-1} shows. However, for the convenience of addressing the challenge "Which", which will be explained in next subsection, the sender is not moving directly approaching the receiver, it moves orthogonally to the camera orientation, therefore the complementary angle of $\theta$ which is $\alpha$ is the required angle as Fig. \ref{subfig::3-2} shows. The $\alpha$ can be calculated as
\begin{equation}
\begin{aligned}
\theta\ &=\ \arccos(\frac{c}{v_S}(1-\frac{f_0}{f}))\\
\alpha\ &=\ |\frac{\pi}{2}-\theta|.
\end{aligned}
\label{eqt::theta}
\end{equation}
As long as $|\alpha|<FOV/2$, the user with the receiver is in the picture.

\begin{figure}[!t]
\centering
\includegraphics[width=0.3\textwidth]{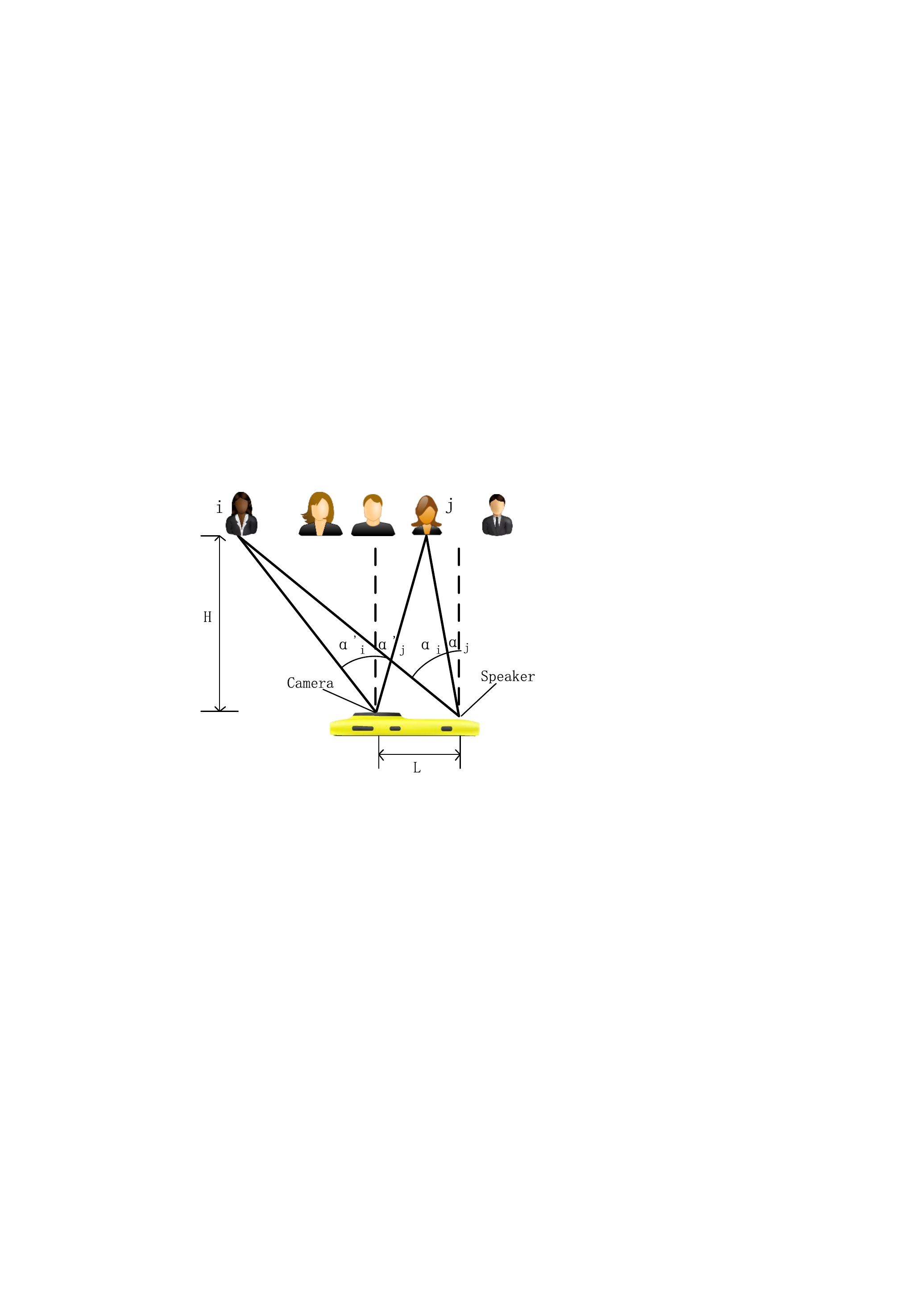}
\caption{The gap between the speaker and the camera. $L$ is the distance of the gap, $H$ is the distance from the sender to the receiver, $\alpha_i$ and $\alpha_j$ are the angles from the receiver i and j to the speaker, and the $\alpha_i'$ and $\alpha_j'$are the angle which is required from the receiver i and j to the camera.}
\label{fig::Gap4}
\end{figure}

However, there is a gap between the speaker and the camera, as Fig. \ref{fig::Gap4} shows. The angle $\alpha$ is from each receiver to the the speaker of the sender, and the required one is the angle to the camera, then the gap may cause unexpected angular error of determining who are in the picture. Suppose the gap distance between the camera and the speaker is $L$, the distance between the sender and the receivers is $H$, $\alpha$ is the angle of the receiver to the sender speaker, $\alpha'$ is the angle of the receiver to the sender camera, then the angular error $|\alpha-\alpha'|$ can be calculated according to
\begin{equation}
\alpha'\ =\ \arctan (\tan \alpha \pm L/H).
\label{eqt::angularerror}
\end{equation}
Table \ref{tab::angularerror} lists the angular errors at three common photographing distances, we use the gap distance of Sumsung S4 as $L$, which is around 95mm. We can see that the maximum angular error is around $1^\circ$, if we assume that the average shoulder width of a 18-year-old female is 35cm\cite{ShoulderWidth}, the angular difference of two persons standing next to each other at a 3m distance to the camera is around $6.65^\circ$ which is far larger than the angular error. Hence, the gap between the camera and the speaker would not affect the result of judging a person whether she is in a picture or not.

\begin{table}[!hbp]
\centering
\caption{Angular errors due to the gap between camera and speaker}
\label{tab::angularerror}
\begin{tabular}{|c|c|c|c|c|c|c|c|}
\hline
\backslashbox{Distance}{$\alpha$} & $0^\circ$ & $10^\circ$ & $20^\circ$ & $30^\circ$ & $40^\circ$ & $50^\circ$ & $60^\circ$\\
\hline
3m & $1.8^\circ$ & $1.7^\circ$ & $1.5^\circ$ & $1.3^\circ$ & $1.0^\circ$ & $0.7^\circ$ & $0.4^\circ$\\
\hline
5m & $1.0^\circ$ & $1.0^\circ$ & $0.9^\circ$ & $0.8^\circ$ & $0.6^\circ$ & $0.4^\circ$ & $0.2^\circ$\\
\hline
10m & $0.5^\circ$ & $0.5^\circ$ & $0.4^\circ$ & $0.4^\circ$ & $0.3^\circ$ & $0.2^\circ$ & $0.1^\circ$\\
\hline
\end{tabular}
\end{table}

\begin{figure}[!t]
\centering
\includegraphics[width=0.45\textwidth]{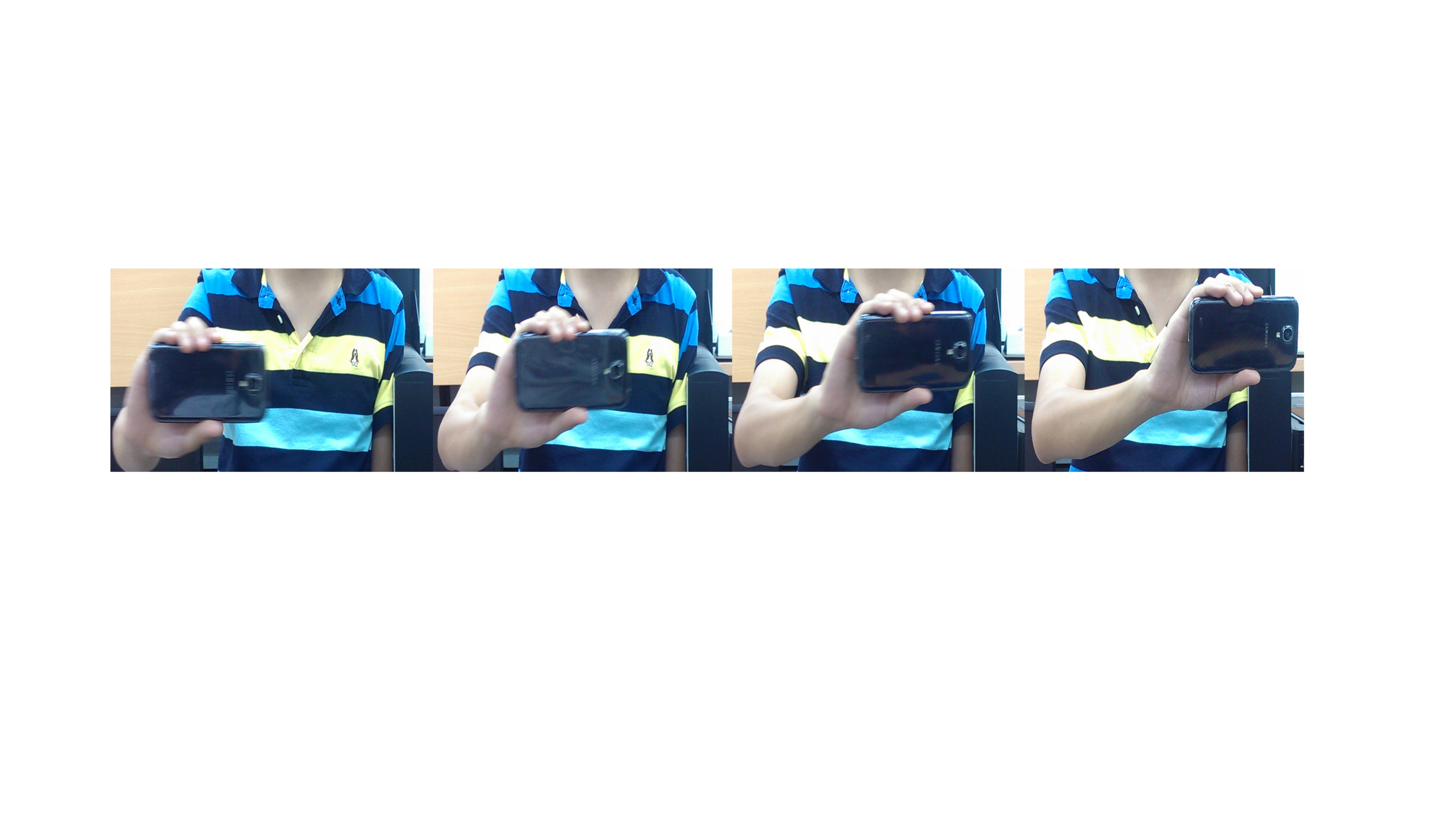}
\caption{Scanning gesture of simulating the movement of speaker.}
\label{fig::scanning}
\end{figure}

According to Eqt. \eqref{eqt::theta}, the velocity $v_S$ of the sender speaker is known beforehand. However, the movement of the speaker is simulated by the hand gesture in this paper, as Fig. \ref{fig::scanning} shows, we have to measure the velocity and take the peak velocity as the $v_S$. Most off-the-shelf mobile phones are embedded with Accelerometer which gives three dimensional accelerations of the device movement. As the way the mobile phone is moved, we use the readings of Y-axis as the acceleration value of $a$, and calculate the velocity $v$ according to
\begin{equation}
\label{eqt::atov}
v\ =\ v_0\ +\ \sum\limits_{t=1}^n{a(t)\cdot \Delta t},
\end{equation}
where $v_0$ is the initial velocity (usually equals to 0), $\Delta t$ is the sampling interval which equals to 10ms in this paper, and $n$ is the number of samples. Although the velocity is measured, it may be different from the true value due to the error of the accelerometer, it could bring deviation to the value of $\alpha$. Moreover, the place where the peak velocity appears may not be the place where the camera is using for picturing, it would make another "gap" between the angles to the camera and to the speaker.

\begin{figure}[!t]
\centering
\begin{minipage}[t]{0.22\textwidth}
\centering
\includegraphics[width=\textwidth]{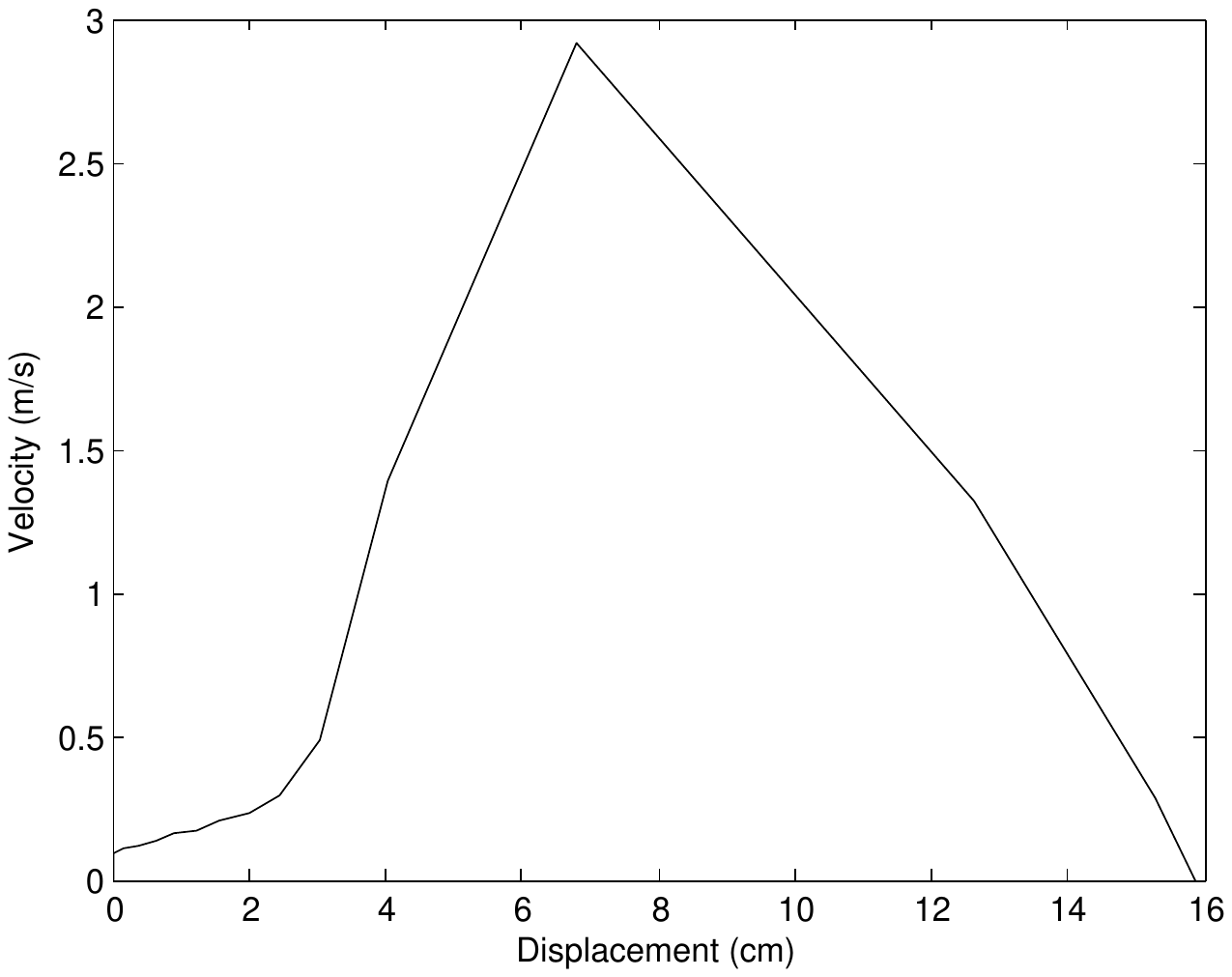}
\caption{Relation between velocity and displacement.}
\label{fig::velocityanddisplacement}
\end{minipage}
\hfil
\begin{minipage}[t]{0.22\textwidth}
\centering
\includegraphics[width=\textwidth]{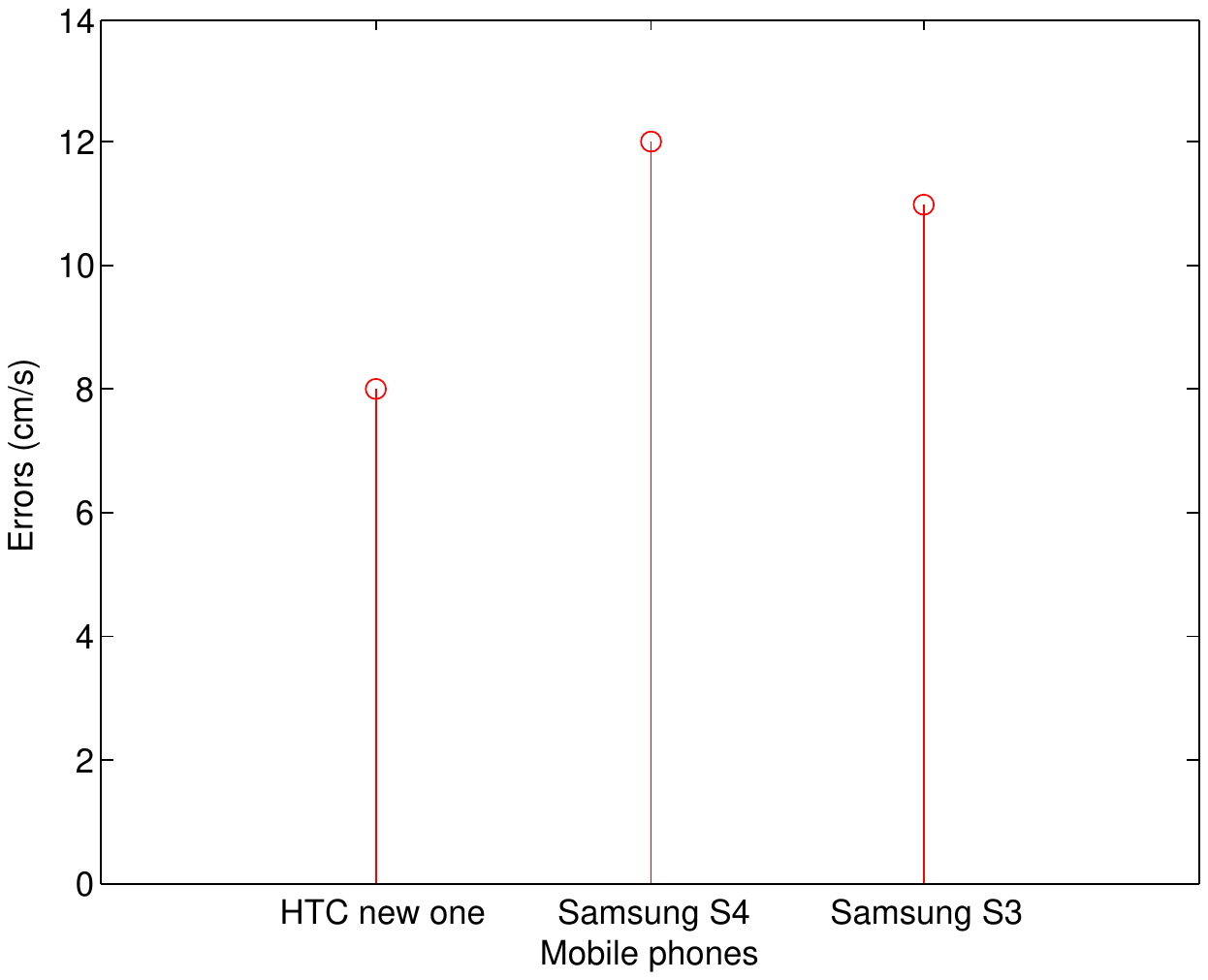}
\caption{Velocity errors of the accelerometer.}
\label{fig::errors}
\end{minipage}
\end{figure}

As Fig. \ref{fig::velocityanddisplacement} shows, the peak velocity appears in the middle of the whole displacement of the mobile phone approximately, therefore, through presetting the displacement trajectory of the mobile phone and put the camera in the middle point for picturing could eliminate the mentioned "gap".

To test the error of the accelerometer, we use four different mobile phones, including two HTC new one, Samsung S4 and Samsung S3. We move the mobile phone directly towards the target device, and the target device will give the received frequency $f$. According to Eqt. \eqref{eqt::theta}, $c$ equals to 340m/s, $f_0$ equals to 20KHz, $f$ is detected by the target device and $\theta$ equals to 0, then the true value of $v_S$ can be calculated. Fig. \ref{fig::errors} shows the error of $v_S$ of three different mobile phones, we can see that the average error is around 10 cm/s which would make $1^\circ$ deviation according to Eqt. \eqref{eqt::theta}.

Because different mobile phone has different gap between the speaker and the camera and the error of the accelerometer, it would be difficult to correct the errors. However, the evaluation still shows good results despite of the impact of the errors.

% How to "Which"? the resolution? how about multi-rows? how to determine the relative position?
\subsection{Which are you in the picture?}

We consider two positioning scenarios: One row positioning and Multi-row positioning. One row positioning is usual in real life photography when attendees are around three or four. Multi-row positioning is common in group photography when there are too many attendees to line up in one row. We will discuss how to identify each person in the picture so that she can be tagged accordingly in the two scenarios.

\noindent\textbf{Single row localization}

Once the audio frame omitted by the sender arrives at a receiver, the receiver will apply the FFT transform on it, and the frequency observed by the receiver can be calculated as $f\ =\ \frac{c}{c-v_S\cos\theta}f_0\cdot \frac{N_{FFT}}{F_s}$, where $\theta$ is the angle from the receiver to the moving direction of sender speaker, $N_{FFT}$ is the number of FFT points, and $F_s$ is the sampling rate. To correctly identify the relative position of each receiver, the frequency shifts $\Delta f$, which equals to $f-f_0$, should at least one FFT point different from the other $\Delta f$s. Assume the angle $\theta$ of receiver A is $\alpha$, of receiver B positioning next to receiver A is $\beta$, and $\cos\alpha\ >\ \cos\beta$, then the requirement can be denoted as
\begin{equation}
\label{eqt::alphabeta}
\begin{aligned}
\left(\frac{1}{c-v_S\cos\alpha}-\frac{1}{c-v_S\cos\beta}\right)\frac{c\cdot f_0 \cdot N_{FFT}}{F_s}\ >\ 1.
\end{aligned}
\end{equation}

Suppose the $\alpha$ is known in advance, then the $\beta$ can be described through this inequality
\begin{equation}
\label{eqt::beta}
\begin{aligned}
\cos\beta\ <\ \frac{-Q\cdot c + v_S\cdot\cos\alpha\cdot(1+Q)}{v_S\cdot(1-Q+Q\cdot v_S \cdot\cos\alpha/c)},
\end{aligned}
\end{equation}
where $Q\ =\ \frac{F_s}{f_0\cdot N_{FFT}}$. Because the function $\cos x$ is monotonically decreasing when $x\in(0,\pi)$, the lower limit of $\beta$ is then obtained. We list a number of possible values of $\alpha$ (the common FOV is around $70^\circ$, so the $alpha$ usually ranges from $55^\circ$ to $125^\circ$) as Table \ref{tab::bminusa44100} shows, to see the minimum required value of $\beta$ where $f_0\ =\ 20KHz$, $F_s\ =\ 44100Hz$, $N_{FFT}\ =\ 2048$ and $v_S\ =\ 3.4\ m/s$.

\begin{table}[!h]
\centering
\begin{minipage}{0.2\textwidth}
\centering
\caption{Angular Resolution before Undersampling}
\begin{tabular}{c c c}
\Xhline{1.2pt}
$\alpha$ & $\beta$ & $\beta - \alpha$\\
\hline
55.0 & 62.1 & 7.1\\
65.0 & 71.6 & 6.6\\
75.0 & 81.3 & 6.3\\
85.0 & 91.2 & 6.2\\
95.0 & 101.2 & 6.2\\
105.0 & 111.5 & 6.5\\
115.0 & 122.1 & 7.1\\
125.0 & 133.0 & 8.0\\
\hline
\end{tabular}
\label{tab::bminusa44100}
\end{minipage}
\begin{minipage}{0.2\textwidth}
\centering
\caption{Angular Resolution after Undersampling}
\begin{tabular}{c c c}
\Xhline{1.2pt}
$\alpha$ & $\beta$ & $\beta - \alpha$\\
\hline
55.0 & 56.1 & 1.1\\
65.0 & 65.9 & 0.9\\
75.0 & 75.9 & 0.9\\
85.0 & 85.8 & 0.8\\
95.0 & 95.8 & 0.8\\
105.0 & 105.9 & 0.9\\
115.0 & 115.9 & 0.9\\
125.0 & 126.0 & 1.0\\
\hline
\end{tabular}
\label{tab::bminusa6300}
\end{minipage}
\end{table}
% figure 5
\begin{figure*}[!t]
\centerline{
\subfloat[]{\includegraphics[width=0.4\textwidth]{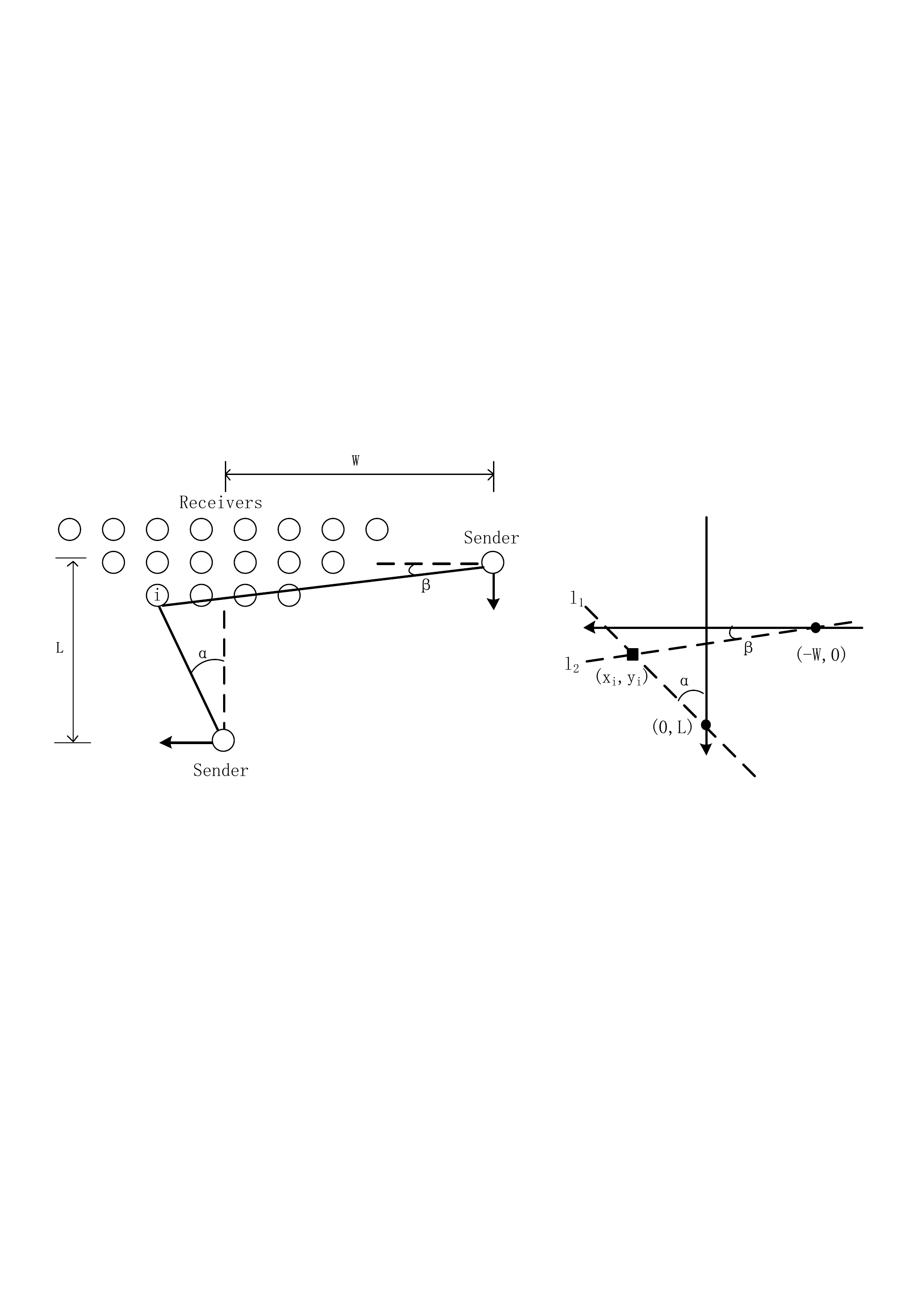}
\label{subfig:multirowdemon}}
\hfil
\subfloat[]{\includegraphics[width=0.3\textwidth]{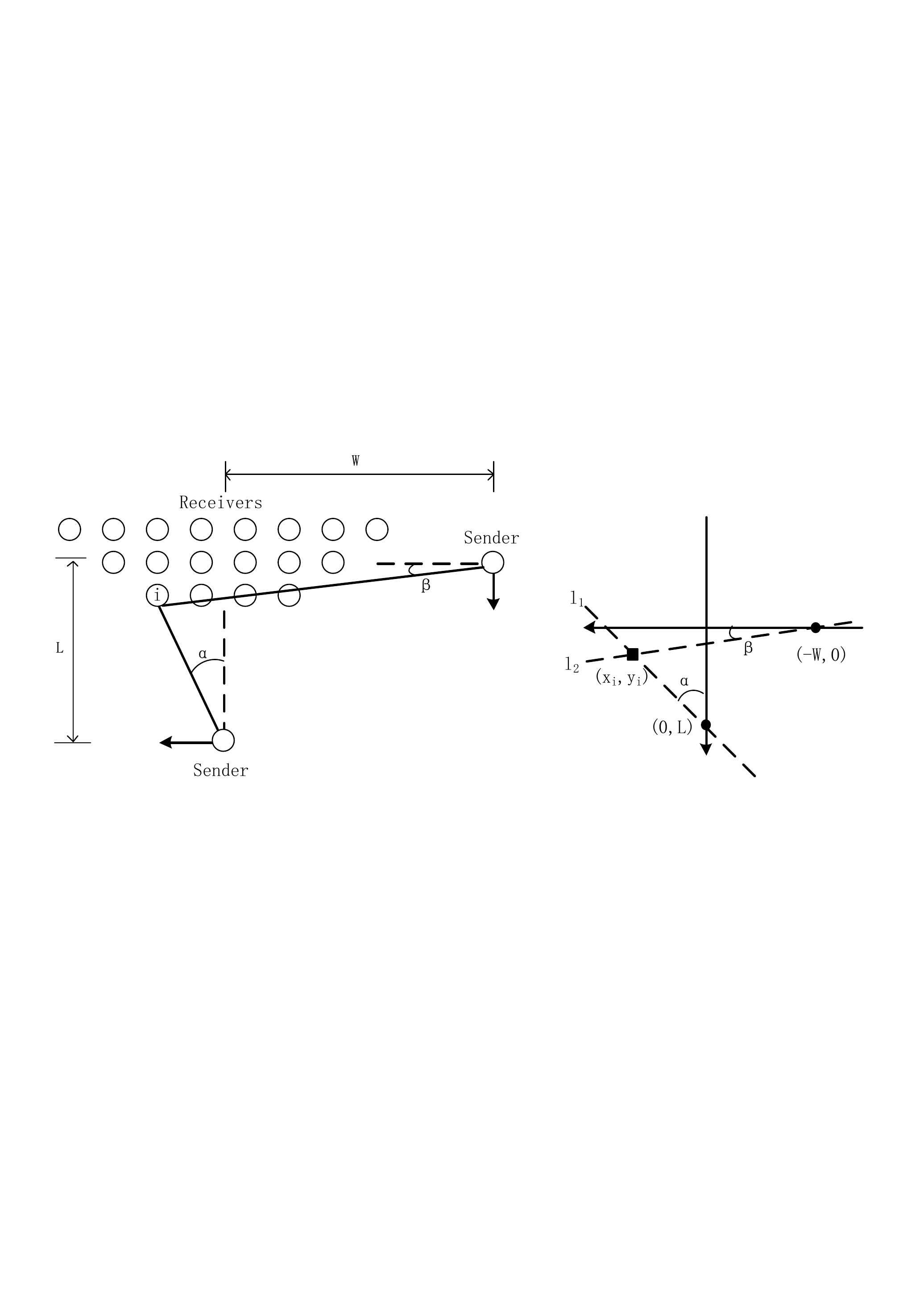}
\label{subfig:multirowxy}}
}
\caption{Illustration of multi-row localization. (a) demonstrates that the sender will choose tow location for moving the speaker, each location is W and L distance from the center point of the receiver group respectively. (b) maps the relative locations of receiver i and sender to a Cartesian coordinate system, to demonstrate how to calculate the value of $(x_i,y_i)$. }
\label{fig:multirows}
\end{figure*}

We can see that the minimum required angular resolution is at least $8.0^\circ$ which exceeds $6.65^\circ$ (the angle between two 18-year-old girls standing next to each other at a 3m distance to the camera). To reduce the lower limit of $beta$ and the requirement of angular resolution, the variable Q in Eqt. \eqref{eqt::beta} should be tuned down because $\beta$ is proportional to the Q. There are three options to tune down the Q: (1) increasing the sending frequency $f_0$, (2) increasing the number of FFT points $N_{FFT}$, (3) decreasing the sampling rate $F_s$. For a common mobile phone, the up limit of frequency range is lower than 22KHz, which limits the improvement of angular resolution by increasing the $f_0$.

The second and third options are basically identical to increasing the frequency resolution. In fact, the only way to increase the frequency resolution is to increase the time length $T$ with the signal\cite{FFTzero}. Because $T=\frac{N_{FFT}}{F_s}$, either $N_{FFT}$ should be increased or $F_s$ should be decreased. Increasing the FFT point would involve higher computational cost which is not suitable for the energy-constrained mobile phone. The left way is to decrease the sampling rate $F_s$ and fix the $N_{FFT}$. The increment of $T$ in the receiver end will prolong the response time, however, the system is not strictly demanding in time and it would be acceptable.

Given that, if the bandwidth of a bandpassing signal is significantly smaller than the central frequency of the signal, it is possible to sample the signal at a much lower rate than the Nyquist sampling rate without causing the alias\cite{Spartacus}. This technique is called Undersampling. The undersampling technique could be described as follows: assume the lowest and the highest band limits of the audio tone is $f_L$ and $f_H$, respectively. According to the undersampling theorem, the condition for an acceptable new sampling rate is that shifts of the bands from $f_L$ to $f_H$ and from $-f_H$ to $-f_L$ must not overlap when shifted by all integer multiples of the new sampling rate $F_s^*$\cite{undersample}. This condition can be interpreted as the following constraint:
\begin{equation*}
\frac{2\cdot f_H}{n}\leq F_s^* \leq \frac{2\cdot f_L}{n-1},\ \forall :\ 1\leq n \leq \lfloor\frac{f_H}{f_H-f_L}\rfloor,
\end{equation*}
where $n=F_s/F_s^*$, $\lfloor\cdot\rfloor$ is the flooring operation. We use the settings of the Spartacus\cite{Spartacus} that $n=7$, $F_s^*=6.3KHz$, the bandwidth is 2KHz to avoid spectrum aliasing. Using the new sampling rate $F_s^*$, we obtain a new list of $\beta$ values, as Table \ref{tab::bminusa6300} shows. We can see that the angular resolution is reduced to around $1^\circ$ which is sufficient to distinguish two neighboring persons.

Now that the angular resolution is sufficient to distinguish the receivers, the sender can sort the frequency shifts of all receivers to get the relative positions. We assume that the speaker always moves toward the left direction, so the receiver with the largest frequency shifts is the first person from left to right, and the receiver with the smallest frequency shifts (the frequency shifts could be negative) is the last person on the right.

\noindent\textbf{Multi-row localization}

The above discussion is assuming that the receivers are positioning in one single row, if the receivers are forming multiple rows, the relative position is no longer described in one dimension. We use a two dimensional coordinate system to denote the relative position of each receiver. Assume $(x_i,y_i)$ denotes the relative position of receiver i, $x_i$ is the relative position in one row, $y_i$ is the row number of receiver i. The question is how to acquire the value of $y_i$.

The intuitive idea is to leverage the signal strength, because different row has different distance to the sender and has different received signal strength. However, the signal strength is sensitive to obstacles, the variation is significant for the same receiver when the mobile device is held in hand or placed in the pocket. Furthermore, it would be difficult to determine that the difference of the received signal strength is caused by different rows or different positions in the same row.

Since the movement of the speaker can be used to determine the value of $x_i$ in one single row, it is possible to get the two dimensional value $(x_i,y_i)$ through two-time movements of the speaker. As Fig. \ref{subfig:multirowdemon} shows, the receivers forms three rows, the speaker of the sender needs two movements in two different locations A and B to obtain the value of $(x_i,y_i)$. Each time the camera of the sender faces to a point of the receiver group (usually the point is the center point), and the speaker moves in a constant velocity $v_0$ towards the left. The distance from the camera to the center of the group is L and W in position A and B respectively. The angle from receiver i to the camera in position A is $\alpha$ and in the position B is $\beta$, then the angles can be obtained through the Doppler effect formulation.

Now we will explain how to determine the value of $(x_i,y_i)$ through two-time movements of the sender speaker. Fig. \ref{subfig:multirowdemon} can be mapped to a Cartesian coordinate system as Fig. \ref{subfig:multirowxy} shows. Certainly the receiver i is the intersection of two lines $l_1$ and $l_2$. As we already know the slope, x-intercept and y-intercept of both lines, we can obtain the intersection point $(x_i,y_i)$ according to
\begin{equation}
\label{eqt::xy}
\begin{aligned}
&l_1:\quad y-L\ =\ \tan(\alpha+\pi/2)\cdot x\\
&l_2:\quad y\ =\ \tan(\beta)\cdot (x+W)\\
\\
&\tan(\alpha+\pi/2)\cdot x + L = \tan(\beta)\cdot (x+W)\\
&\Rightarrow \left\{
\begin{split}
&x\ =\ \frac{\tan(\beta)\cdot k - 1}{\tan(\alpha+\pi/2)+\tan(\beta)}\cdot L\\
&y\ =\ (\frac{\tan(\alpha+\pi/2)\cdot(\tan(\beta)\cdot k - 1)}{\tan(\alpha+\pi/2)+\tan(\beta)}+1)\cdot L,\\
\end{split}
\right.
\end{aligned}
\end{equation}
where $k=W/L$ is an adjustable parameter in calculating the value of $(x_i,y_i)$. However, the value of $y_i$ is not the row number yet, we have to cluster all the receivers into rows according to the value of $y_i$.

Spectral clustering is one of the most popular modern clustering algorithm, it is simple to implement, and can be running efficiently on mobile devices\cite{SpectralClustering}. The spectral clustering algorithm was proposed to solve the Graph Cut problem in the graph theory, it is based on a similarity graph which needs to transform the values of $y_i$ into a weighted adjacent matrix. Since the value of $(x_i,y_i)$ is the relative position of each receiver, we consider to represent the adjacent relationship by transforming the values of $y_i$ according to Eqt. \eqref{eqt::W} in the Euclidean space.
\begin{equation}
\label{eqt::W}
W_{ij}\ =\ e^{-(y_i-y_j)^2}
\end{equation}
After getting the adjacent matrix W, we can calculate the diagonal degree matrix D as Eqt. \eqref{eqt::D} and the unnormalizad Laplacian matrix L by Eqt. \eqref{eqt::L}.
\begin{equation}
\label{eqt::D}
D_{ii}\ =\ \sum\limits^j{W_{ij}}
\end{equation}
\begin{equation}
\label{eqt::L}
L\ =\ D\ -\ W
\end{equation}
After that, the k smallest eigen values and corresponding eigen vectors of L will be obtained by Singular Value Decomposition (SVD) algorithm\cite{SVD}. At last the matrix formed by the eigen vectors will be clustered by k-means algorithm\cite{kmeans} where the final clustering result is given.

Assume $R_i$ is the average value of $y_i$ of cluster i, it can be calculate as
\begin{equation}
\label{eqt::Ri}
R_k\ =\ \sum\limits_{y_i\in\Omega_k}^{n_k}{y_i}/{n_k},
\end{equation}
where $\Omega_k$ is the kth cluster of $y_i$, $n_k$ is the number of $y_i$ in $\Omega_k$. By sorting $R_i$, we can get the order of rows. The smallest $R_i$ refers to the first row, and the largest one refers to the last row.

\section{Implementation and Evaluation}\label{sec::evaluation}

To validate the scheme, we implemented a system on the Android platform on various mobile phones, including HTC new one, Samsung S4, and Samsung S3. We test the performance of "Who" and "Which" of the system in different scenarios and compare with an existing vision based tagging system Picasa to evaluate the system.

\subsection{System Design}

\begin{figure}[!t]
\centering
\includegraphics[width=0.4\textwidth]{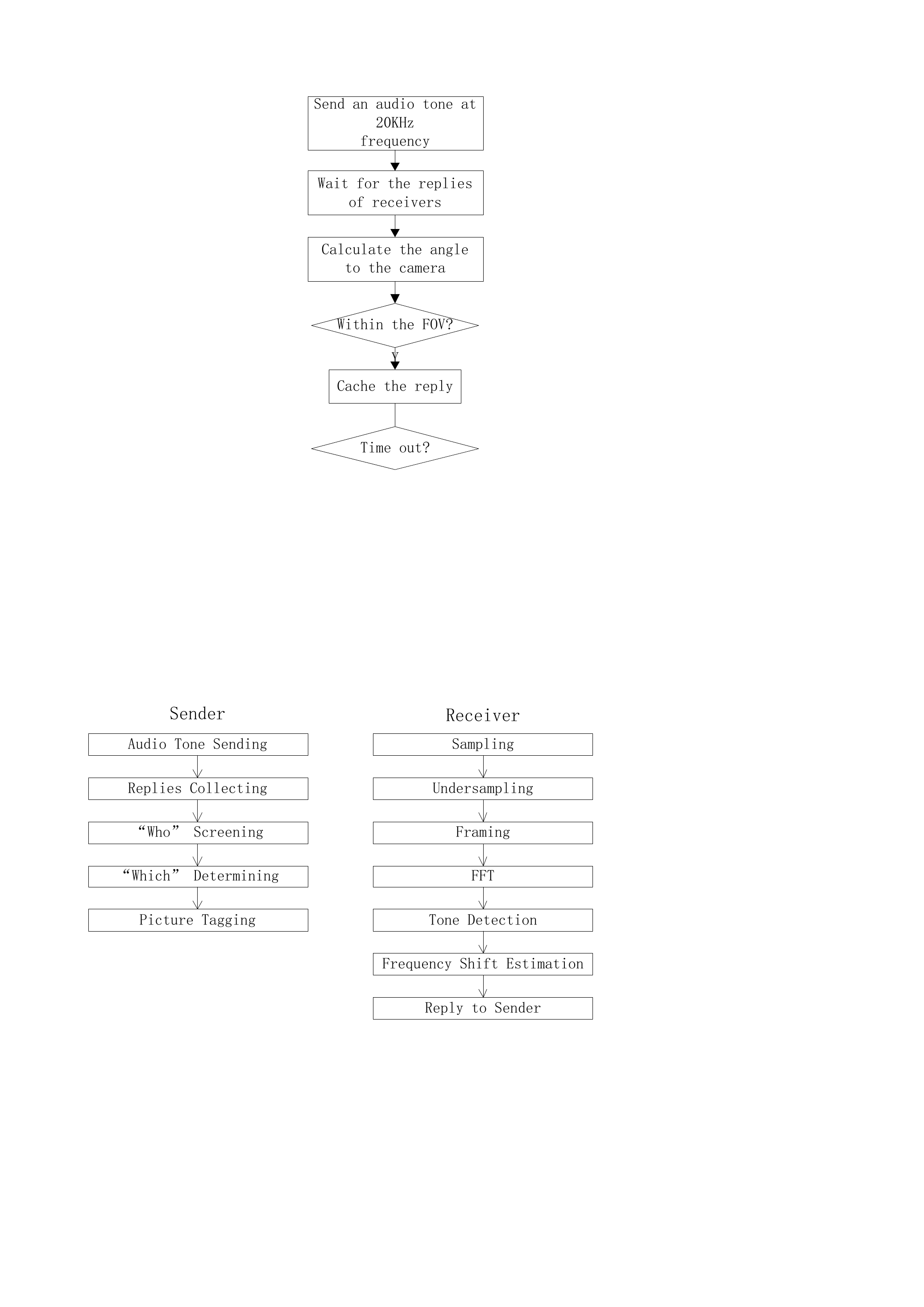}
\caption{Flowchart of the sender and receiver.}
\label{fig::flowchart}
\end{figure}

Fig. \ref{fig::flowchart} shows an overview of the system. We will describe the system in two parts: sender and receiver.

\subsubsection{Sender}

Before taking pictures, Alice will create a private group to let her friends to join in so that the irrelevant people will not appear in the image tags. However, not all the group members will appear in each picture, it still needs to screen the ones who are not in the picture.

Alice's mobile device, as the sender, will send a command to each receiver to activate its microphone for the next coming audio tone through WiFi Ad-hoc network or carrier's network, then it will send an audio tone at 20KHz frequency (because it is inaudible for humans\cite{Audio20KHz}) with 2KHz bandwidth for 1 second. During the audio emitting process, Alice moves her phone in a scanning gesture to the left direction. As mentioned before, if Alice's friends positioned in more than one row, Alice would have to move to another location, and move the mobile phone again.

After audio tone emitting, the sender will wait for the reply of each receiver until time out. The reply consists of the user name and the frequency shifts detected by the corresponding receiver.

According to Eqt. \eqref{eqt::theta}, the sender can obtain the angle $\alpha$ of each receiver, if $f$, $f_0$, $c$ and $v_S$ is known. $f$ can be calculated through $f=f_0+\Delta f$, where $\Delta f$ is the frequency shifts received from the receiver. $f_0$ equals to 20KHz, $c$ is a constant value 340m/s, however, $v_S$ is unknown because the movement of the speaker is simulated by hand gesture. As mentioned in section \ref{sec::methodology}, the velocity can be calculated by the embedded accelerometer according to Eqt. \eqref{eqt::atov}, and the peak velocity is used as $v_S$ for calculating the angle $\alpha$. So that the one has $|\alpha| > FOV/2$ will be screened out.

If there is only one row of receivers, the frequency shifts will be sorted in a descending order indicating the relative position of each person in the picture from left to right, and the picture will be tagged with user names accordingly. If there are multiple rows of receivers, the pair value $(x_i,y_i)$ of each receiver i will be calculated by Eqt. \eqref{eqt::xy}, and the spectral clustering process will be performed to classify each receiver to its belonging row. For each cluster, the relative position will be obtained by sorting the $x_i$, and the row order can be calculated by sorting the mean value of $y_i$ of each cluster. Then the picture will be tagged accordingly.

\subsubsection{Receiver}

Alice's friends will join the group Alice established firstly, and their mobile phones, as the receivers, will wait for the activation command from the sender. When the command arrives, it will activate the microphone and begin to listen to the upcoming audio tone.

In the receiver end, we take the pipeline of \cite{Spartacus} as a reference. Once the audio tone is detected, it will be sampled in a 44.1KHz rate (a common sampling rate for commodity audio component) to ensure the 20KHz frequency band can be captured according to the Nyquist theorem. In the undersampling step, a 10-order Butterworth bandpass filter is used to sieve a frequency band centered at 20KHz with 2KHz bandwidth. The undersampling process will preserve every 7th sample to form a new sample set without losing the frequency domain.

As mentioned before, by decreasing the sampling rate and fixing the FFT points, the sampling time will increase and the frequency resolution will increase as a consequence. If we do FFT for the whole undersampled data set, we will get a spectral band consists various frequency components including the shifted frequencies and the unshifted frequency 1.58KHz (undersampled from 20KHz). It would be difficult to detect the boundary line of the shifted frequencies. \cite{SoundWave} proposed to find the boundary line by scanning the frequency bins on both sides of 20KHz independently until the amplitude drops below 10\% of the peak value. However, it is a coarse method to detect the frequency shifts. Because the peak velocity of hand movement usually lasts for around 10ms, if we divide the data set into 10ms frames, the spectrum in each frame would be small enough for detecting the frequency shifts. So, after undersampling, the sampled data set will be divided into 10ms a frame, and each frame has a 75\% overlapping ratio. For each frame, FFT will be performed using 6.3KHz sampling rate and 2048 FFT points. Since 10ms frame contains less than 2048 data points, the FFT will do zero padding to ensure the resolution, although it is not the real resolution. The tone will then be detected by comparing the energy, if the average energy in the frequency range 1KHz from 1.08KHz to 2.08KHz (corresponding to 19.5KHz and 20.5KHz) is over 1.5 times larger than the average energy of the whole frame, we consider the frame contains the frequency band from 19.5KHz to 20.5KHz, and the frequency correlated to the peak energy will be detected as a shifted frequency. However, due to the sampling clock offset (SCO) between the sender and the receiver, the received frequency may not be the original one from the sender, therefore we take the frequency in the first frame (the sender is not moving) as the original frequency $f_0$, and adjust the frequency range in the energy detection accordingly. By detecting the shifted frequencies in all the frames and substract by $f_0$, we take the largest absolute value as the frequency shifts $\Delta f$. However, the resolution of the frequency shifts is obtained by zero padding of FFT, we need to detect the frequency shifts in a real resolution. So, the FFT is performed for the whole undersampled data set using 2048 FFT points, the $\Delta f$ is located in the spectrum and do a border line search as \cite{SoundWave} in a range $f_0+\Delta f\pm \xi$ to find the final $\Delta f$, where $\xi$ is a adjustable parameter which equals to 10Hz in this paper.

After frequency shift extraction, the receiver will reply the frequency shift and the user name to the sender through WiFi Ad-hoc network or carrier's network.

\subsection{Evaluation}

% experiment set up

We implement the system on the Android platform, and conduct experiments in real-life scenarios with 7 mobile phones which are all embedded with a microphone, a speaker and a camera, including HTC new one, Samsung S4 and Samsung S3. One mobile phone is used for picturing, the other 6 ones are the receivers. We test the system in three different locations: student cubicle area, hallway and student activity square, the first two locations are testing how distance and irrelevant people affect the accuracy, and the last location is for testing how ambient noise impact the system function. In each location, six receivers will be positioned to one single row and multiple rows to test the localization performance.
% Spartacus experiment
\subsubsection{Localization Accuracy}

\begin{figure}[!t]
\centerline{
\subfloat[]{\includegraphics[width=0.2\textwidth]{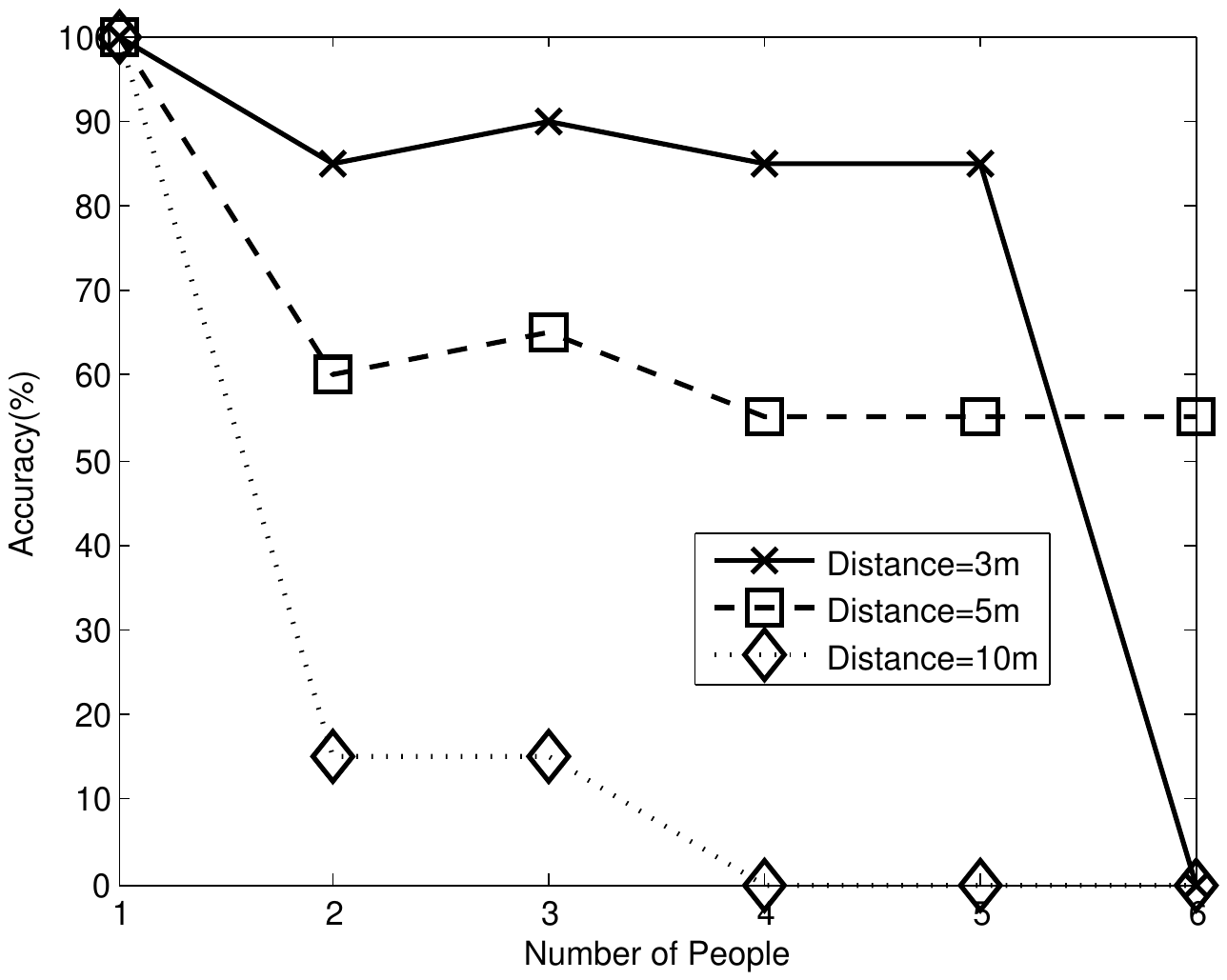}
\label{subfig:accsingle}}
\hfil
\subfloat[]{\includegraphics[width=0.2\textwidth]{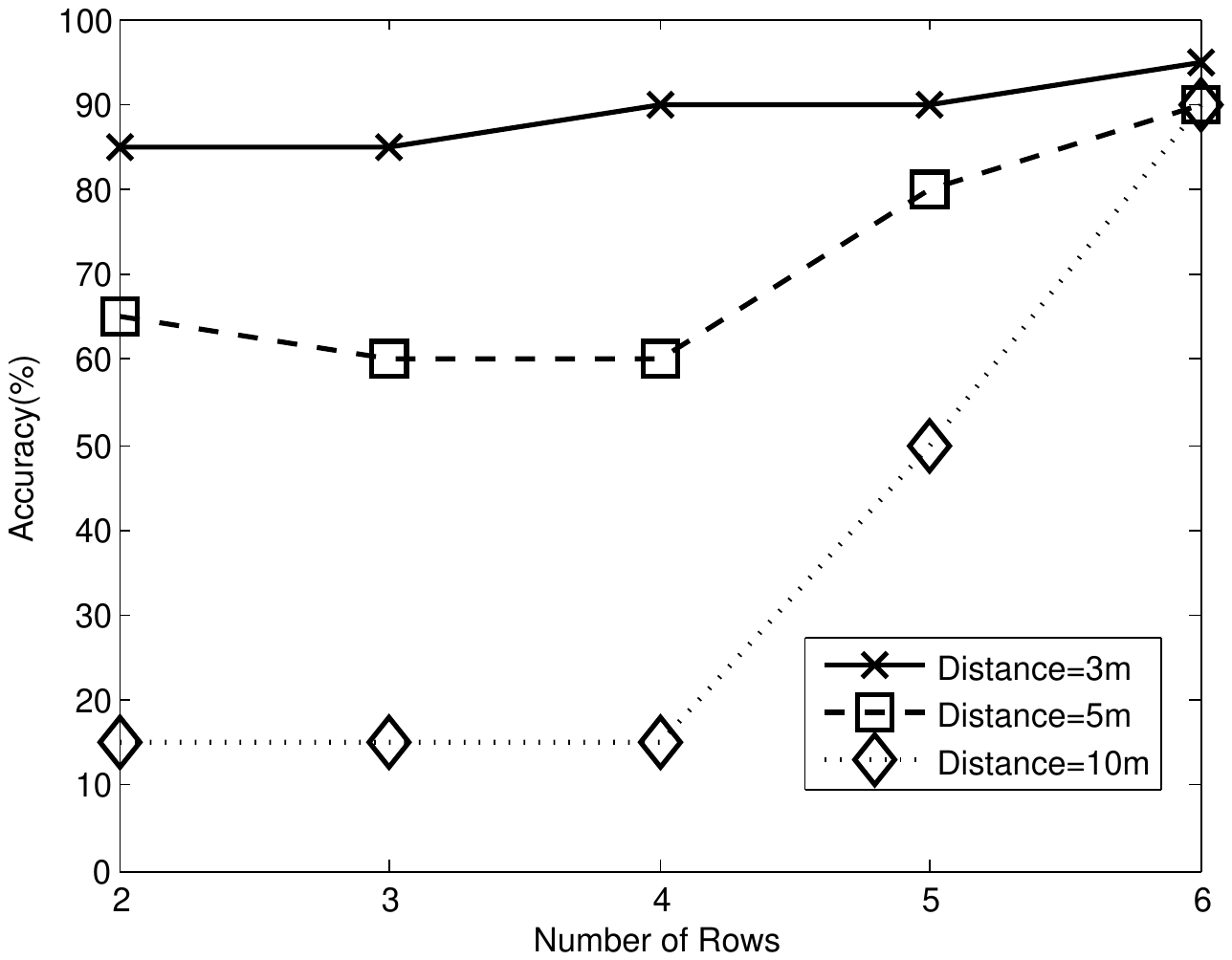}
\label{subfig:accmulti}}
}
\caption{Performance of "Which" at different distances. (a) is the accuracy of localization in one single row. (b) is the accuracy of localization in multiple rows.}
\label{fig:accuracy}
\end{figure}

Because the audio tone would experience severe attenuation as distance grows, we set the volume of audio tone to the maximum 0 dbm and test the performance of "Which". The tests start by picturing one person, and adding one person for each time. Every test will repeat 20 times and the accuracy will be calculated by $accuracy=matched\ times/total\ times$. As Fig. \ref{subfig:accsingle} shows, in a single row scenario, within 3m, the accuracy of localization for the number of receivers below six is above 85\%, the exception is that when the number is larger than six, one of the receivers fall out of the FOV so that the accuracy is affected. Within 5m, the accuracy is decreased to around 55\% and within 10m the signal strength is severely degraded and the accuracy is less than 10\%.

In the multi-row scenario, the first movement will be executed at three different locations while the second movement will be perform 1m away from the group. As Fig. \ref{subfig:accmulti} shows, when the row number is less than three, the accuracy at three distances is similar to the single row scenario, the error seems not accumulated by the second movement of the speaker. When the row number increases, the accuracy also increases, which seems not related to the distance of the first movement. Two reasons may be attributed to this phenomenon: (1) there is only 1m away from the receivers of the second movement which will have higher accuracy. When the number of rows grows, the number of people in each row decreases, the accuracy tends to more related to the second movement. (2) the second movement is mainly for clustering receivers to rows which is somehow error-tolerant.

\begin{figure}[!t]
\centering
\includegraphics[width=0.3\textwidth]{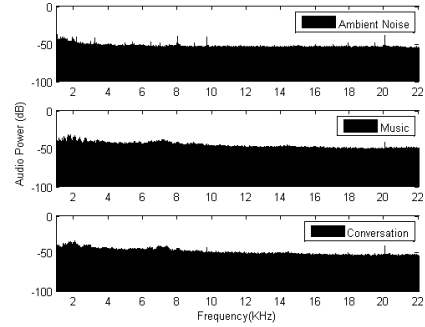}
\caption{Spectrums in three noise scenarios where the speaker is 3m away sending an audio tone at 20KHz.}
\label{fig::environment}
\end{figure}

We also test the performance of the system under ambient noise, music and conversation to see whether the system may work properly in most real-life scenarios. As Fig. \ref{fig::environment} shows, when a speaker emits a 20KHz frequency audio tone at a 3m distance, ambient noise would hardly affect the SNR of the high frequency band, the music and conversation noises somehow decrease the SNR of the high frequency band which may affect the accuracy when SNR is smaller that 10db..

\begin{figure*}[!t]
\centerline{
\subfloat[]{\includegraphics[width=0.2\textwidth]{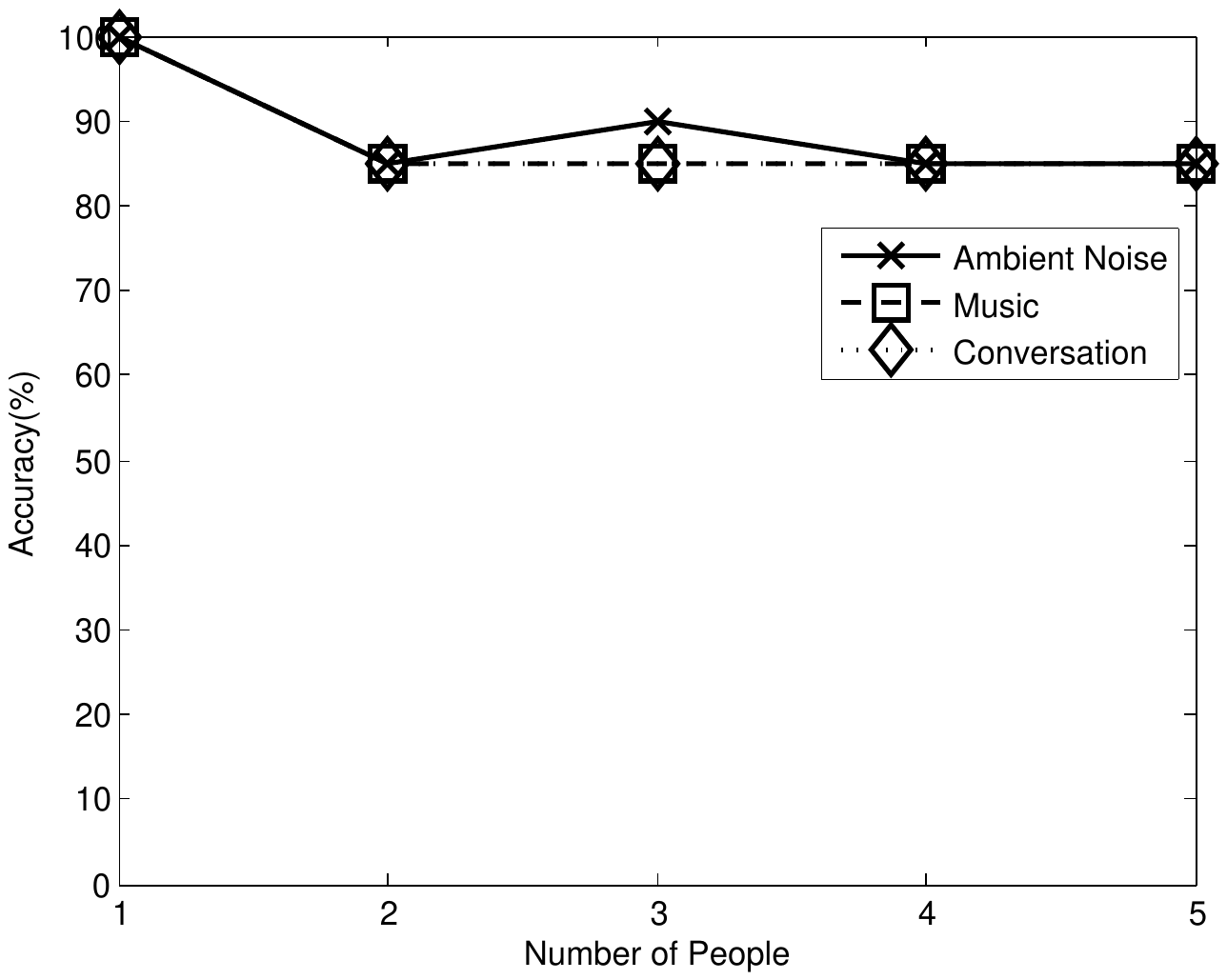}
\label{subfig:3maccsingle}}
\hfil
\subfloat[]{\includegraphics[width=0.2\textwidth]{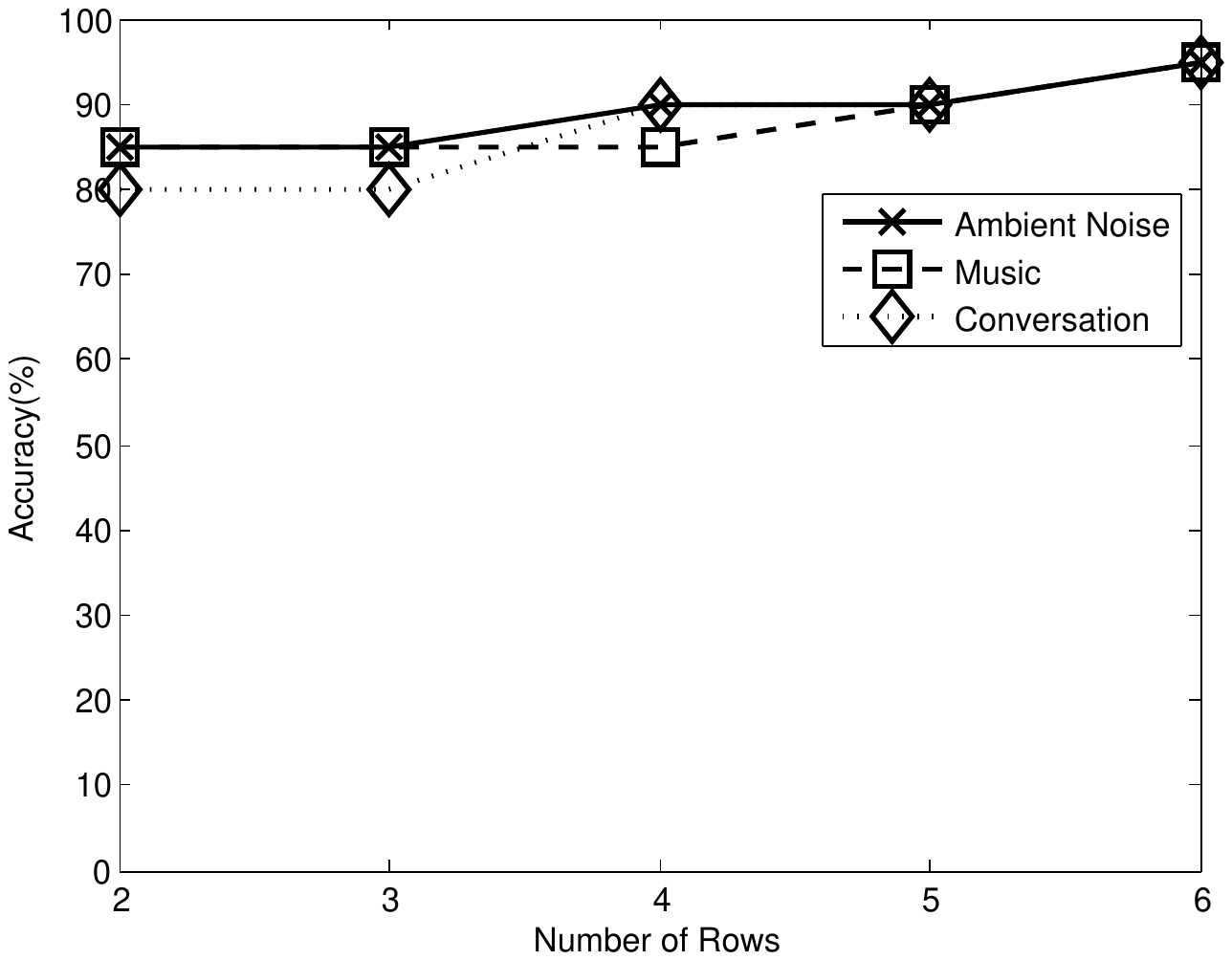}
\label{subfig:3maccmulti}}
\hfil
\subfloat[]{\includegraphics[width=0.2\textwidth]{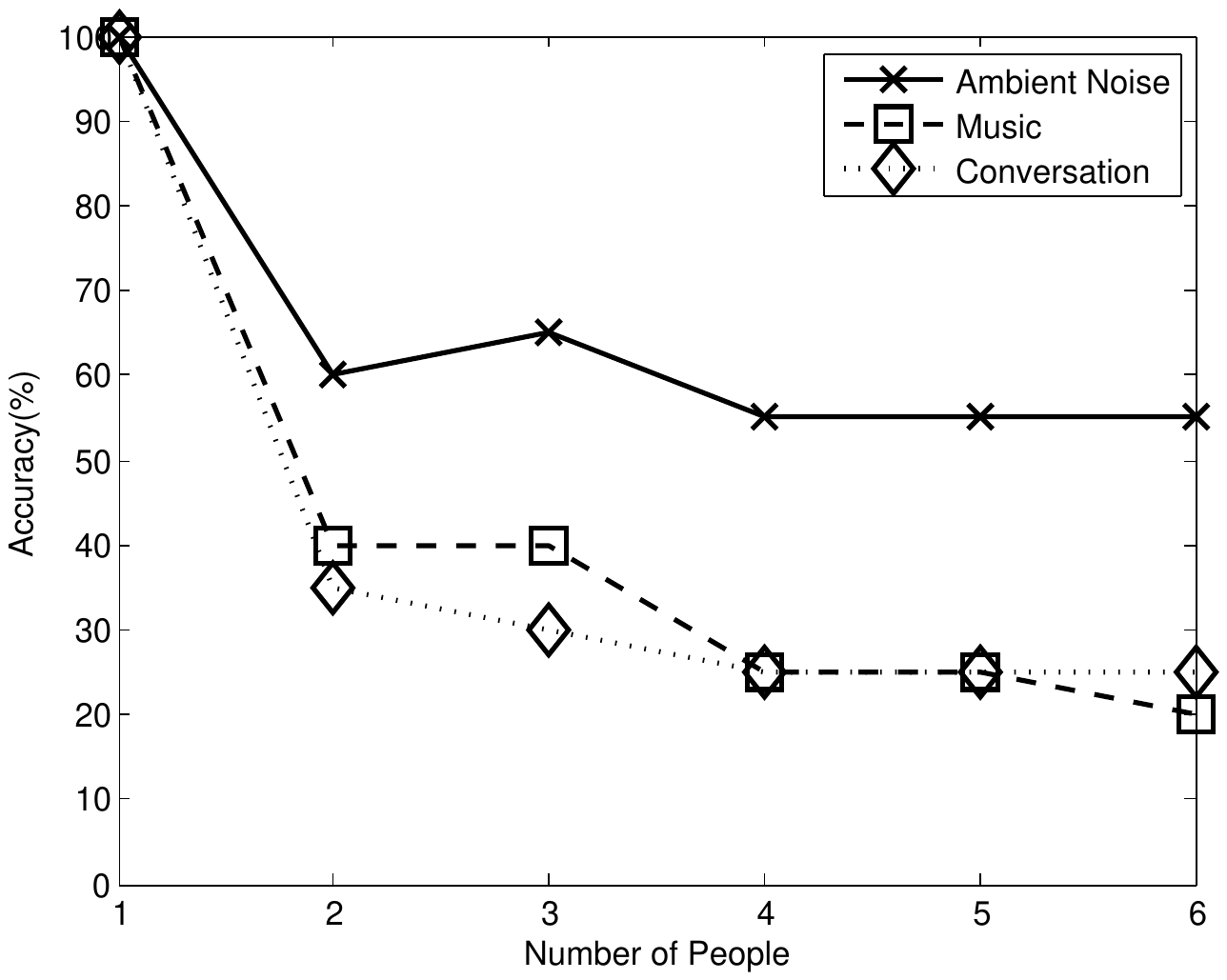}
\label{subfig:5maccsingle}}
\hfil
\subfloat[]{\includegraphics[width=0.2\textwidth]{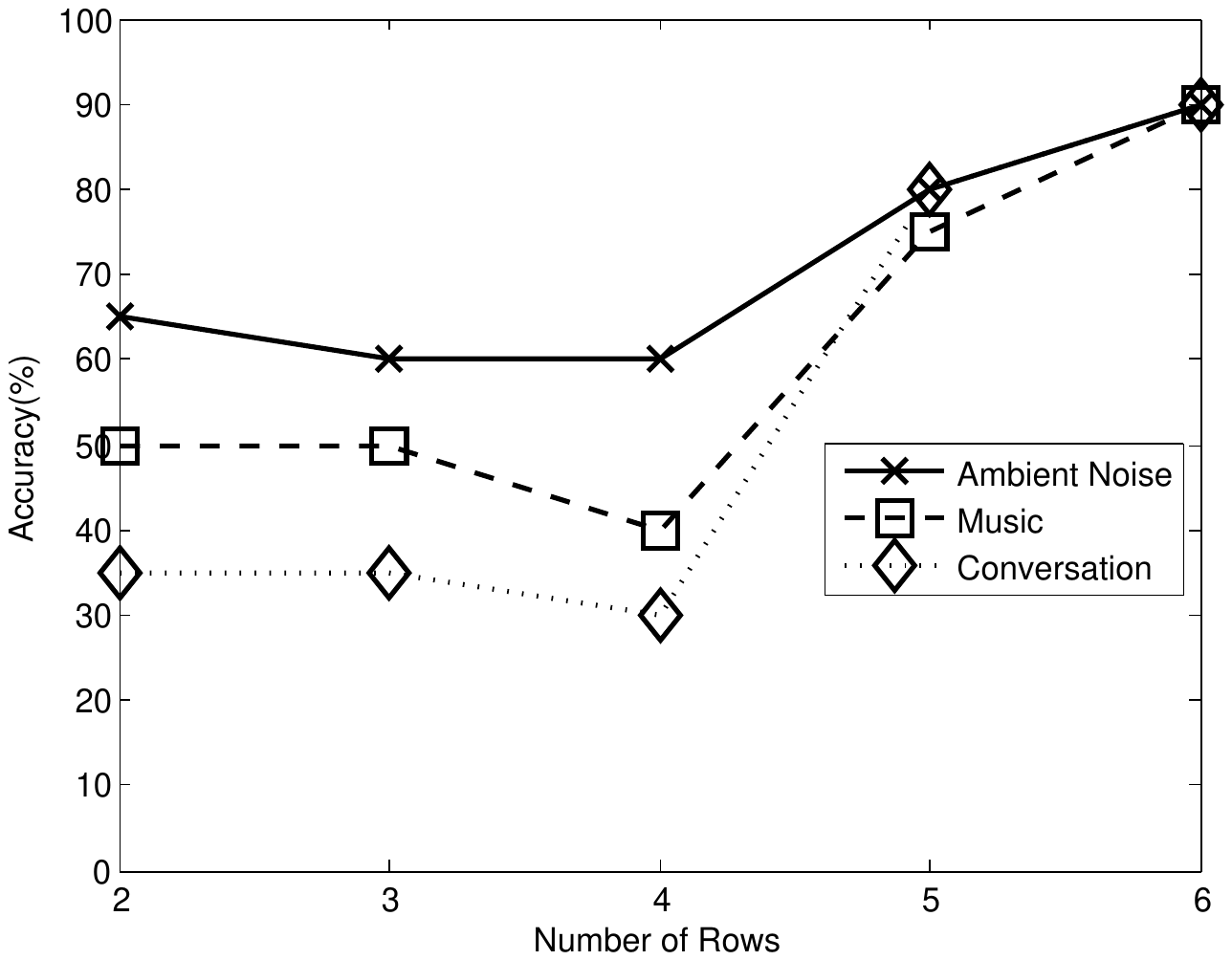}
\label{subfig:5maccmulti}}
}
\caption{Performance of "Which" at different noise scenarios. (a) is the accuracy in one single row within 3m. (b) is the accuracy in multiple rows within 3m. (c) is the accuracy in one single row within 5m. (d) is the accuracy in multiple rows within 5m.}
\label{fig:accuracynoise}
\end{figure*}

Fig. \ref{subfig:3maccsingle} and \ref{subfig:3maccmulti} show that at a distance of 3m, the three kinds of noise hardly affect the accuracy. When the distance grows to 5m, the accuracy decreased severely in the music and conversation environment as Fig. \ref{subfig:5maccsingle} and \ref{subfig:5maccmulti} show.

We can see from the experiments that within 3m, the accuracy of single row and multi-row both are above 85\% and are seldom affected by the noises. With the growth of distance, the accuracy falls monotonically and the noises would severely affect the performance. Although 3 meters are sufficient for picturing in many real-life scenarios, it may still needs a high power speaker for longer distance picturing, such as class photography.
% Tagsense experiment: scenes and metric. scenes: single line, two lines with aligned(front line crouch, second line stand), two lines all stand, three lines with first line crouch and other two lines stand. Metrics includes: recall, precision and fallout.
\subsubsection{Tagging Accuracy}

\begin{figure}[!t]
\centering
\subfloat[]{\includegraphics[width=0.35\textwidth]{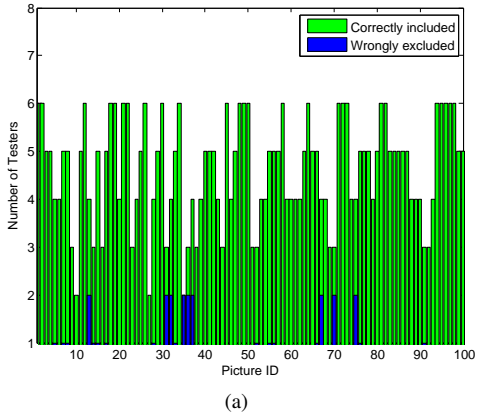}
\label{subfig:include}}
\hfil
\subfloat[]{\includegraphics[width=0.35\textwidth]{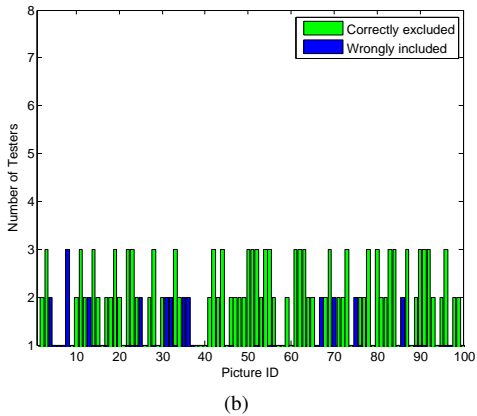}
\label{subfig:exclude}}
\caption{Performance of "Who". (a) and (b) shows correctly include and exclude the testers and wrongly include and exclude the testers.}
\label{fig:correctwho}
\end{figure}

To verify the accuracy of "Who", we have taken 100 pictures in various situations, including different ways of positioning (one row, two row), different number of relevant people (up to six) and irrelevant people (up to three irrelevant people near, far away and in the picture). As Fig. \ref{fig:correctwho} shows, we test the accuracy of how many people correctly be included and excluded. We can see that the system performs well in including relevant people and excluding irrelevant people for most of time.

\begin{figure}[!t]
\centering
\includegraphics[width=0.3\textwidth]{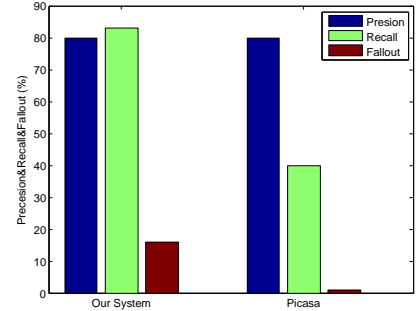}
\caption{The presion, recall and fallout comparison with Picasa.}
\label{fig::precision}
\end{figure}

Also, to see how well the system performs comparing to the existing systems, we take several pictures of our testers for face recognition of Picasa\cite{PicasaFace} to perform name tagging, and we use the following metrics to measure the tagging accuracy.
$$precsion=\frac{|People Inside\cap Tagged|}{|Tagged|}\nonumber$$
$$recall=\frac{|People Inside\cap Tagged|}{|People Inside|}\nonumber$$
$$fallout=\frac{|People Outside\cap Tagged|}{|People Outside|}\nonumber$$
As Fig. \ref{fig::precision} shows, we can see that although the precision and the fallout of Picasa look well, the recall rate which is a key metric for search-like applications is much lower than ours. A low recall implies that when a user searches for a picture, the results are unlikely to include the one she is looking for\cite{TagsenseTransaction}. The balanced performance in three metrics proves that our scheme could perform well in image retrieval area.

Through all the experiments, we can see that the system could perform a robust and high accuracy of device localization within 3m, when the distance grows, a larger power speaker would be necessary or the photographer needs to step forward to move her mobile device firstly and then step backward for picturing, of which the user experience would not be pleasing.

\section{Related Works}\label{sec::relatedworks}

% mobile device selection and localization could enhance social activity, because a user could select a neighboring user and interact with her
As a well-known physical feature of wireless signal, the Doppler effect is widely used in gesture recognition\cite{SoundWave,WholeHomeGesture,Gesture1,Gesture2,Gesture3,Spartacus}, device selection and localization\cite{Doplink,Localization1,Localization2} areas in recent works. \cite{SoundWave} leverages the Doppler effect to recognize user gestures before the computer, the user could move her hands up, down, forward and backward for different commands to the computer, such as page up, page down, volume up and volume down. \cite{WholeHomeGesture} explored the possibility to use the Doppler effect in the WiFi environment.

Obviously, the Doppler effect is not the only solution of device selection and localization, for example, to the same end, Point\&Connect\cite{PointandConnect} accomplished that a user makes a pointing gesture towards the target device to select and interact with it, just as Spartacus\cite{Spartacus} did. The difference is the Point\&Connect measures the variation of distance between the sender and the multiple receivers via BeepBeep\cite{BeepBeep}, only the receiver on the pointing direction observes greatest variation. However, it requires the displacement of the mobile phone to be at least 20cm for an acceptable accuracy. Another example is that BeepBeep\cite{BeepBeep} is a localization method based on radiating sound waves to obtain the distance between the sender and the receiver through calculating the delay from sending the sound wave to receiving the echo. In fact, the localization methods are able to identify the location of each user, and address the mentioned challenges as a consequence. However, the accuracy of most localization methods is too coarse to identify two neighboring users standing next to each other. GPS based methods have around 7 meters accuracy\cite{GPS} and are not available in indoor environment. WiFi based methods have better accuracy of about 2 meters in indoor environment\cite{WiFi} which are still not sufficient. The acoustic TOA (time of arrival) based methods like BeepBeep\cite{BeepBeep} has a very high accuracy of 3cm for localization, it would be possible to leverage the BeepBeep to identify each user's location. Nevertheless, it needs more than one anchor nodes which may not be possible in real-life picturing scenarios.

\section{Conclusion}\label{sec::conclusion}

Automatic image tagging plays an important role in image retrieval and social network area. We propose an alternative way instead of using image recognition technique to leverage the common mobile device and the well-known physical law - the Doppler effect for tagging the pictures while they are taking. Also the relative position of each people in the picture, no matter in one row or multiple rows, can be recognized, therefore the names of each people can be tagged accordingly. As a proof of concept, we implement a system and evaluate it in various real-life scenarios, the results show that the accuracy of relative position recognizing is above 85\% within 3m and the system has balanced performance in precision, recall and fallout rate of picture tagging, which would perform well in the image retrieval area.


%\section*{Acknowledgment}

\bibliographystyle{IEEEtran}
\bibliography{references}
\balance

\end{document}